\documentclass[aps,prd,amsmath,floats,onecolumn,floatfix,nofootinbib,showpacs]{revtex4}
\usepackage{amssymb,graphicx} 
\usepackage{bm}
\usepackage{mathrsfs}
\usepackage[usenames]{color}
\usepackage[normalem]{ulem}

\newcommand{\Mirr}{M_{\text{irr}}}

\newcommand{\ScriPlus}{\mathscr{I}^+}

\begin{document}

\title{Bondi-Sachs energy-momentum for the constant mean extrinsic
  curvature initial value problem}

\author{James M. Bardeen$^1$} \author{Luisa T. Buchman$^2$}
\affiliation{${}^1$Physics Department, University of Washington,
  Seattle, Washington 98195 USA}
\affiliation{${}^2$Theoretical Astrophysics, California Institute of
  Technology, Pasadena, California 91125 USA}

\date{\today}

\begin{abstract}
  The constraints on the asymptotic behavior of the conformal factor
  and conformal extrinsic curvature imposed by the initial value
  equations of general relativity on constant mean extrinsic curvature
  (CMC) hypersurfaces are analyzed in detail.  We derive explicit
  formulas for the Bondi-Sachs energy and momentum in terms of
  coefficients of asymptotic expansions on CMC hypersurfaces near
  future null infinity.  Precise numerical results for the Bondi-Sachs
  energy, momentum, and angular momentum are used to interpret
  physically Bowen-York initial data on conformally flat CMC 
  hypersurfaces similar to that calculated earlier by
  Buchman et al.~\cite{Buchman:2009ew}.
\end{abstract}

\maketitle


\section{Introduction}
\label{Intro}
Recently, there has been increased interest in formulations for
numerical relativity based on conformal
compactification~\cite{FrauendienerLRR} in which the calculational
grid extends to future null infinity ($\ScriPlus$), where the
gravitational radiation amplitude can be read off unambiguously, with
at most numerical errors, and where no dynamical boundary conditions
are necessary.  In principle this can be done with Cauchy
characteristic matching methods, but these have not been implemented
successfully with non-trivial dynamics.  Cauchy characteristic
extraction methods can extrapolate from the outer boundary of a
conventional Cauchy code to determine waveforms at $\ScriPlus$, but do
not eliminate errors in the Cauchy development deriving from inexact
boundary conditions at a finite radius.  The characteristic methods
are reviewed by J. Winicour~\cite{Winicour2009}.  Our focus here is
the initial value problem on hyperboloidal spacelike hypersurfaces,
and specifically the case of constant mean extrinsic curvature (CMC)
hypersurfaces~\cite{Husa2006,Moncrief2009,
  Rinne2009,Zenginoglu2010,Bardeen2011a}.  The vanishing of the
conformal factor $\Omega$ at $\ScriPlus$ accounts for the singular
behavior of the physical spacetime metric at $\ScriPlus$ in
compactified coordinates.  The conformal geometry, in which
$\ScriPlus$ is an ingoing null hypersurface, is regular in a
neighborhood of $\ScriPlus$, and a CMC hypersurface of the physical
spacetime is spacelike in the conformal spacetime out to and including
its intersection with $\ScriPlus$, a 2-surface with spherical topology
which we denote by $\dot{\ScriPlus}$.  Regularity conditions relating
the 2D extrinsic curvature of $\dot{\ScriPlus}$ as embedded in the CMC
hypersurface to the conformal extrinsic curvature need to be satisfied
at $\dot{\ScriPlus}$, but when imposed in the initial data are
automatically preserved by the evolution
equations~\cite{Moncrief2009,Bardeen2011a}.  The constraint equations,
given suitable gauge conditions, determine the leading behavior of the
conformal factor and the conformal extrinsic curvature of the
hypersurface in the neighborhood of $\ScriPlus$ in terms of asymptotic
gravitational wave amplitudes~\cite{Andersson1994}.  If the physical
mean extrinsic curvature is not too large, the dynamics of the sources
(black holes, neutron stars) takes place where the CMC hypersurfaces
are not that different from the hypersurfaces of conventional 3+1
methods.

Conformally flat initial data on CMC hypersurfaces
were considered in Ref.~\cite{Buchman:2009ew}.  The conformal 
momentum constraint equations have the same form as they do on 
maximal hypersurfaces, and with conformal flatness admit a class 
of analytic solutions which are slight generalizations of the 
well-known Bowen-York solutions~\cite{Bowen-York:1980} often 
used for single or multiple black hole initial data on maximal 
hypersurfaces.  The Hamiltonian constraint provides an
elliptic equation for the conformal factor which is degenerate at 
$\ScriPlus$.  Despite this degeneracy, Ref.~\cite{Buchman:2009ew} 
obtained numerical solutions without much difficulty using the standard
spectral elliptic solver of the Caltech-Cornell-CITA {\tt SpEC}
code~\cite{SpECwebsite,Pfeiffer2003}.  The degeneracy constrains 
the leading terms in the expansion of $\Omega$ about $\dot{\ScriPlus}$.

An essential part of the physical interpretation of these solutions is
knowing precisely the total energy, linear momentum, and angular
momentum of the system as coded in the asymptotic behavior of the
spacetime metric at $\ScriPlus$.  The standard Arnowitt-Deser-Misner
(ADM) formulas~\cite{ADM} for these quantities only apply on
asymptotically flat slices at spatial infinity.  The Bondi-Sachs
energy and momentum at $\ScriPlus$~\cite{Bondi1962,Sachs1962} are the
relevant quantities for CMC hypersurfaces.  Ref.~\cite{Buchman:2009ew}
did not fully address this issue, making only some rather crude
estimates of the total energy, momentum, and angular
momentum with limited validity and for the most part with uncertain
errors.

While there is an extensive literature dealing with the problem of
extracting these global physical quantities near or at at future null infinity
(see, e.g., the review~\cite{Szabados2004}), part of which deals 
specifically with CMC hypersurfaces~\cite{Chrusciel2004}, we see
practical difficulties in implementing many of these procedures.  Some
of them require the choice of a reference spatial metric and extrinsic
curvature.  Determining the reference quantities with the precision
necessary to obtain unambiguous results requires considerable 
effort, in general.  Formulas based on the asymptotic
behavior of the Weyl tensor (e.g.,~\cite{Sachs1962,
Penrose1963,Stewart1989}) can be useful computationally, but an
analysis based on geometrically defined spatial coordinates gives 
more insight into the asymptotic geometry of CMC hypersurfaces and 
its relationship to the Bondi expansion on null hypersurfaces.

We first consider the problem of calculating the Bondi energy and 
momentum in general asymptotically flat spacetimes foliated by 
CMC hypersurfaces.  Following Appendix A of Ref.~\cite{Bardeen2011a}, 
we analyze the asymptotic geometry using Gaussian normal spatial 
coordinates tied to the $\dot{\ScriPlus}$ 2-surface.   
This gauge choice allows a relatively simple characterization of the
asymptotic behavior of the spacetime metric and extrinsic
curvature on a CMC hypersurface based on asymptotic solutions 
of the constraint equations.  
Constructing the asymptotic coordinate transformation from the 
CMC-based coordinates to Bondi-Sachs null coordinates gives us 
the Bondi-Sachs mass aspect in terms of geometric objects
defined on a single CMC hypersurface.  The monopole and dipole moments
of the Bondi-Sachs mass aspect are the Bondi-Sachs energy and linear
momentum, respectively.  In the simple case of conformally flat
initial data, the mass aspect is just a sum of contributions
from the asymptotic expansions of the conformal factor and conformal
extrinsic curvature.

The Bowen-York solutions for the conformal extrinsic curvature 
contain parameters such as boost and spin
vectors, which on maximal hypersurfaces are simply related to the ADM
momentum and angular momentum.  On CMC hypersurfaces, the
relationship of the Bowen-York parameters to physical momenta and
energies is more complicated, but we are able to obtain analytic 
expressions for the direct contributions of the
conformal extrinsic curvature to the Bondi-Sachs energy and 
momentum in terms of the Bowen-York parameters.  Unlike on  
maximal hypersurfaces, on conformally flat CMC hypersurfaces the
coordinate displacement of the black hole from the center of the
coordinate sphere representing $\ScriPlus$ enters in a non-trivial
way.  The conformal factor contributions to the Bondi-Sachs energy and
momentum are extracted from the numerical solution of
the Hamiltonian constraint equation.  We also derive an analytic 
expression for the total angular momentum of the system which 
only depends on the Bowen-York parameters, without any 
contribution from the conformal factor.  We discuss the physical 
interpretation of some representative examples of Bowen-York 
initial data on CMC hypersurfaces similar to those of 
Ref.~\cite{Buchman:2009ew}, and show how to construct 
initial data approximating a circular-orbit binary black hole system.

The Wald sign convention for the extrinsic curvature is adopted, so
that the mean extrinsic curvature $K$ is positive for hypersurfaces
with diverging future-directed normals, the hypersurfaces 
extending to future null infinity rather than to spatial infinity ($K =0$) 
or to past null infinity ($K< 0$).  Our notation is an amalgam of that of
Refs.~\cite{Buchman:2009ew,Chrusciel2004} and very similar to that in
Appendix A of Ref.~\cite{Bardeen2011a}, but be alert to occasional
deviations.  We assume vacuum in a neighborhood of future null
infinity.  The discussion of the asymptotic geometry on general 
CMC hypersurfaces is in
Sec.~\ref{Sec:2}.  In Sec.~\ref{Sec:3}, we use the lapse and shift
which preserve the CMC hypersurface condition and the Gaussian normal
spatial coordinate condition to facilitate the asymptotic
transformation to Bondi-Sachs coordinates.  The final expression for
the mass aspect depends only on the geometric properties of a single
CMC hypersurface.  Numerical results for the generalized Bowen-York
initial data on conformally flat CMC hypersurfaces are presented in
Sec.~\ref{Sec:4}.  The implications of our results for astrophysically 
interesting initial data are summarized in
Sec.~\ref{Sec:Discussion}.  In Appendix~\ref{AppA}, we derive the
analytic expressions for the conformal extrinsic curvature
contribution to Bondi-Sachs four-momentum and in Appendix~\ref{AppB} 
the analytic expression for the total angular momentum.

\section{Asymptotic behavior on CMC hypersurfaces}
\label{Sec:2}
The fact that the Hamiltonian constraint equation for the conformal
factor and the momentum constraint equation for the extrinsic
curvature are degenerate at $\ScriPlus$ means that the first few terms
in a power series expansion of these quantities away from
$\dot{\ScriPlus}$ are uniquely determined from the expansion of the conformal 
spatial metric, in spite of the elliptic character of the equations and
therefore the generally nonlocal character of the solutions.  In our
derivation of formulas for the Bondi-Sachs energy-momentum 4-vector,
we will not assume the conformal spatial metric is flat, though
all of our numerical examples are for the conformally flat case.  The
expansion parameter defined in Ref.~\cite{Buchman:2009ew} is only
meaningful for a flat conformal spatial metric, so the analysis here
follows the derivation of asymptotic behavior presented in Appendix A
of Ref.~\cite{Bardeen2011a}, which in turn rather closely follows that
of Ref.~\cite{Andersson1994}.  It is based on a particular choice of
spatial coordinates, Gaussian normal coordinates, in which the
``radial'' coordinate $z$ is the conformal proper distance inward from
$\dot{\ScriPlus}$ along normal spatial geodesics in the CMC
hypersurface considered as a slice of the conformal spacetime.  
As in Ref.~\cite{Bardeen2011a}, we assume that the
intrinsic geometry of $\dot{\ScriPlus}$ is that of a 2-sphere of fixed
area in the conformal geometry, which can be enforced in any
asymptotically flat spacetime by suitable boundary conditions on the
initial conformal gauge and its subsequent evolution, at least if the
gauge evolution variables are determined by elliptic equations, as in
Refs.~\cite{Moncrief2009,Bardeen2011a}.  Angular coordinates $x^A$ are
propagated along the normal geodesics starting from standard polar
angles on the $\dot{\ScriPlus}$ two-sphere.  While the actual
computational coordinates will be different, in general, the
coefficients in the expansions are expressed in terms of quantities
defined covariantly and therefore can be computed in any coordinate
system.  An equality which holds only at $\ScriPlus$ is indicated by a
dot above the equal sign.

We just give the key definitions and results here, and refer the
reader to Ref.~\cite{Bardeen2011a} for details of the derivations.
The conformal spacetime metric $\tilde{g}_{\mu \nu}$ is related to the
physical spacetime metric $g_{\mu \nu}$ by
\[
    \tilde{g}_{\mu \nu} = \Omega^2 g_{\mu \nu},
\]
and a tilde will be used generally to distinguish quantities
associated with the conformal, rather than the physical, geometry.
The conformal factor is required to vanish on
$\ScriPlus$, where it cancels the infinities in the physical metric
due to the compactified spatial coordinates and the CMC hypersurfaces
becoming asymptotically null, and is strictly positive in the
interior.  The general form of the conformal spatial metric in
Gaussian normal coordinates is
\begin{equation}
    \tilde{g}_{ij} dx^i dx^j = dz^2 + \tilde{h}_{AB} dx^A dx^B,
\end{equation}
with $z = 0$ at $\dot{\ScriPlus}$.  A metric function $\xi$ is defined
such that $\det \tilde{h}_{AB} = \xi^{-4}\sin^2\theta$.  At
$\dot{\ScriPlus}$, $\tilde{h}_{AB} \doteq \xi_0^{-2} \breve{h}_{AB}$,
with $\breve{h}_{AB}$ the standard metric of the unit two-sphere in
polar coordinates.  We require that $\xi_0$ be independent of angle.
Then the 2D scalar curvature of $\dot{\ScriPlus}$ is $2\xi_0^2$.  The
only way to preserve this property of $\dot{\ScriPlus}$ is to demand
that $\partial_t \xi_0 = 0$ during the evolution as a condition on the
evolution of the conformal gauge.  We can keep the CMC time coordinate
$t$ equal to a Minkowski retarded time coordinate $u$ at $\ScriPlus$
by imposing the boundary condition $\tilde{\alpha} \doteq K/(3\xi_0)
\equiv \tilde{\alpha}_0$ on the elliptic equation for the conformal
lapse which follows from $\partial_t K = 0$.

A general form for the expansion of $\tilde{h}_{AB}$ in powers of $z$
away from $\dot{\ScriPlus}$ is then
\begin{equation}
    \tilde{h}_{AB} = \xi^{-2} \left[ \breve{h}_{AB} - 2 \breve{\chi}_{AB} z + 
    \left( \breve{\chi}^{CD} \breve{\chi}_{CD} \breve{h}_{AB} - 
    \breve{\psi}_{AB} \right) z^2 + O(z^3) \right].
    \label{Eq:hexpansion}
\end{equation}
Here $\breve{\chi}_{AB}$ and $\breve{\psi}_{AB}$ are traceless,
symmetric tensors on the unit sphere with indices lowered and raised
by $\breve{h}_{AB}$ and its inverse.  Note the identity that
$\breve{\chi}_A^{\;C}\, \breve{\chi}_{CB} \equiv \frac{1}{2}
\breve{\chi}^{CD} \breve{\chi}_{CD} \breve{h}_{AB}$.  As argued in
Ref.~\cite{Bardeen2011a}, we believe it is consistent and physically
appropriate in the context of gravitational radiation generated by the
internal dynamics of isolated systems not to allow any polyhomogeneous
terms, that is, terms containing powers of $\log{z}$ as well as powers
of $z$, to this order in the expansion.  The expansion of $\tilde{h}_{AB}$ 
is closely related to the expansion of the angular part of the metric in 
Bondi-Sachs coordinates.  Ref.~\cite{Chrusciel1995} showed that the 
coefficient of the leading polyhomogeneous term of order $r^{-1}\log r$ 
or $r^{-2}\log r$ in the Bondi expansion is a constant of the motion, so if 
these terms are absent initially they never appear.  If these polyhomogeous 
terms are not present in the Bondi expansion, the corresponding terms 
are not present in Eq.~\ref{Eq:hexpansion}. The inverse of the angular 
part of the metric is  
\begin{equation}
    \tilde{h}^{AB} = \xi^2 \left[ \breve{h}^{AB} + 2 \breve{\chi}^{AB} z + 
    \left( \breve{\chi}^{CD} \breve{\chi}_{CD} \breve{h}^{AB} + 
    \breve{\psi}^{AB} \right) z^2 + O(z^3) \right].
\end{equation}

The angular metric functions can be related to the extrinsic curvature
$\tilde{\kappa}^A_B$ of the constant-$z$ two-surfaces as embedded in
the 3D conformal geometry of the CMC hypersurface.  The trace of the
2D extrinsic curvature is $\tilde{\kappa} = 2 \partial_z (\log{\xi})$,
and the traceless part is
\[
\hat{\tilde{\kappa}}^A_B = \breve{\chi}^A_B + \breve{\psi}^A_B z +
O(z^2).
\]
The expansion of $\xi$ as a power series in $z$ can be expressed in
terms of the expansion of $\tilde{\kappa}$,
\[
    \tilde{\kappa} = \tilde{\kappa}_0 + \tilde{\kappa}_1 z + 
    \tilde{\kappa}_2 z^2 + O(z^3),
\]
with the result 
\begin{equation}
  \log{\xi} = \log{\xi_0} + \frac{1}{2} \tilde{\kappa}_0 z + 
  \frac{1}{4} \tilde{\kappa}_1 z^2 + \frac{1}{6} \tilde{\kappa}_2 z^3 
  + O(z^4).
    \label{Eq:xiexp}
\end{equation}
Satisfying our condition on the intrinsic geometry of
$\dot{\ScriPlus}$ requires that $\xi_0$ be a constant, independent of
both time and angle, but all of the coefficients in the expansion of
$\tilde{\kappa}$ will in general depend on the time $t$ and the
angular coordinates $x^A$.  The 2D extrinsic curvature is a tensor
which can be calculated in any coordinate system by solving the
geodesic equations in the conformal spatial geometry.

A straightforward calculation gives the conformal 3D Ricci tensor
components.  The result is Eq.~(A6) of Ref.~\cite{Bardeen2011a},
\begin{eqnarray}
    \label{Eq:Conformal3DRicci}
    \tilde R^z_{\,z} &=& \partial _z \tilde \kappa - 
    \hat{\tilde \kappa} _{AB} \hat{\tilde \kappa}^{AB} 
    - \frac{1}{2}\tilde \kappa ^2 , \nonumber \\
    \tilde R^z_{\,A} &=& \frac{1}{2}\tilde \kappa _{|{A}}  
    - \hat{\tilde \kappa}^{C}{}_{{A}|{C}} ,  
    \\ \tilde R^{A}_{B} &=&\,
    \partial_z\hat{\tilde \kappa}^{A}_{B} 
    - \tilde\kappa \,\hat{\tilde \kappa}^{A}_{B} +
     \frac{1}{2}\left( {\,^2 \tilde R + \partial _z \tilde \kappa  - \tilde
     \kappa ^2 } \right)\delta^{A}_{B},
     \nonumber
\end{eqnarray}
with 
\begin{equation}
    \label{Eq:Ricci2D}
    \,^2 \tilde{R}^A_B = \frac{1}{2} \,\,^2 \tilde{R} \,\delta^A_B = 
    \xi_0^2 \left[ 1 + \left( \tilde{\kappa}_0 + \frac{1}{2} 
        \tilde{\kappa}_0{}^{\breve{\scriptscriptstyle{|}}C}{}
        _{\breve{\scriptscriptstyle{|}}C} - \breve{\chi}^{CD}{}_
        {\breve{\scriptscriptstyle{|}}CD} \right) z 
      + O(z^2) \right] \delta^A_B.
\end{equation}
The $\scriptstyle{|}$ symbol denotes a covariant derivative with
respect to the metric $\tilde{h}_{AB}$ of the constant-$z$
two-surface, while the $\breve{\scriptscriptstyle{|}}$ symbol denotes
a covariant derivative on the unit sphere.  The curvature scalar
evaluated at $\dot{\ScriPlus}$ is
\begin{equation}
    \tilde{R} \doteq 2 \xi_0^2 + 2 \tilde{\kappa}_1 - \frac{3}{2} 
    \tilde{\kappa}_0^2 - \breve{\chi}^{AB} \breve{\chi}_{AB}.
\end{equation}

We define the 3D conformal extrinsic curvature $\tilde{K}_{ij}$ to
have the same relationship to the time derivative of the conformal
spatial metric, the conformal lapse, and the shift as the physical
extrinsic curvature $K_{ij}$ has to the time derivative of the
physical spatial metric, the physical lapse, and the shift.  This
corresponds to the convention in Ref.~\cite{Bardeen2011a} and the
usual convention in the literature, but differs from
Ref.~\cite{Buchman:2009ew}.  The conformal extrinsic curvature is
decomposed into its trace $\tilde{K}$ and traceless part
$\hat{\tilde{K}}_{ij}$.  The sign of the extrinsic curvature is that
of Wald, with $K > 0$ on a CMC hypersurface extending to $\ScriPlus$.

The initial value equations on a CMC hypersurface constrain
$\hat{\tilde{K}}_{ij}$ through the momentum constraint.  The
well-known ``zero-shear'' condition necessary for $\ScriPlus$ to exist
as a regular null hypersurface in the conformal manifold, expressed in
our coordinates, says that
\begin{equation}
    \Sigma^A_B \equiv \hat{\tilde{K}}^A_B - \frac{1}{2} \delta^A_B 
    \hat{\tilde{K}}^C_C \doteq \hat{\tilde{\kappa}}^A_B \doteq 
    \breve{\chi}^A_B, \qquad 
    \hat{\tilde{K}}^z_A \doteq \hat{\tilde{K}}^z_z \doteq 0.
\end{equation}
The Hamiltonian constraint equation is an elliptic equation for
$\Omega$, degenerate at $\dot{\ScriPlus}$, where $\Omega \doteq 0$.  
The degeneracy allows no freedom in the first few terms of the expansion 
of the solution in powers of $z$.
The asymptotic solution from Ref.~\cite{Bardeen2011a} is
\begin{equation}
    \label{Eq:AsympOmega}
    \Omega = \frac{K}{3} z \left[ 1 - \frac{1}{4} \tilde{\kappa}_0 z + 
      \frac{1}{6} \left( \xi_0^2 + \frac{1}{4} \tilde{\kappa}_0^2 -
        \tilde{\kappa}_1 - \breve{\chi}^{AB} \breve{\chi}_{AB} \right) 
      z^2 + \left( c_3 + \frac{1}{8} Q \log{\frac{K z}{3}} \right) z^3 
        + O(z^4) \right],
\end{equation}
where $c_3$ is a function of angle only known from the global 
(numerical) solution of the elliptic equation and 
\begin{equation}
    \label{Eq:Qdef}
    Q \equiv \xi_0^2 \left[ \breve{\chi}^{AB}{}_{\breve{\scriptscriptstyle{|}}AB} 
    + \breve{\chi}^A_B \left( \breve{\psi}^B_A - \frac{1}{2} 
    \tilde{\kappa}_0 \breve{\chi}^B_A \right) \right].
\end{equation}
The logarithmic term is present whenever outgoing radiation is present at 
$\ScriPlus$.  It is a consequence of the CMC hypersurface condition.  
The coefficient $Q$ also appears in the asymptotic expansion of the 
conformal extrinsic curvature derived from the momentum constraint 
equation:
\begin{equation}
    \label{Eq:AsympKzz}
    \hat{\tilde{K}}^z_z = \breve{\chi}^A_B \breve{\chi}^B_A z + 
    \left( d_1 - Q \log{\frac{K z}{3}} \right) z^2 + O(z^3).
\end{equation}
If, in addition to the zero-shear condition, we require that the Weyl tensor 
vanish at $\ScriPlus$, the expansion of $\Sigma^A_B$ is 
\begin{equation}
    \Sigma^A_B = \breve{\chi}^A_B + \frac{1}{2} \tilde{\kappa}_0 
    \breve{\chi}^A_B z + O(z^2),
    \label{Eq:zeroWeylCMC}
\end{equation}
and the expansion of $\hat{\tilde{K}}^z_A$ is 
\begin{equation}
    \hat{\tilde{K}}^z_A = \breve{\chi}^B_{A\breve{\scriptscriptstyle{|}}B} z 
    + e_1 z^2 + O(z^3).
\end{equation}
The coefficients $d_1$ and $e_1$ are functions of angle which are
known only from a global solution of the momentum constraint equation.
The vanishing of the Weyl tensor at $\ScriPlus$, called the Penrose
regularity condition in Ref. ~\cite{Bardeen2011a}, is a necessary
condition for the absence of polyhomogeneous terms in the Bondi-Sachs
asymptotic expansion based on special null coordinates (see
Ref.~\cite{Sachs1962}).

The time evolution of the conformal factor $\Omega$ is given by 
\begin{equation}
    \partial_t \Omega = \beta^k \partial_k \Omega - \frac{1}{3} \tilde{\alpha} 
    \left( K - \Omega \tilde{K} \right),
    \label{Eq:Omegaevol}
\end{equation}
from which we see that $\tilde{K}$ controls how the conformal gauge
evolves.  We want the conformal gauge to be consistent with
$\dot{\ScriPlus}$ being a two-sphere at the coordinate position $z =
0$ at all times.  Keeping $\ScriPlus$ at $z = 0$ requires the boundary 
condition $\beta^z \doteq \tilde{\alpha} \doteq \tilde{\alpha}_0$ on the shift,
so $\partial_t \Omega = 0$ at $z = 0$.  The only way to make sure that
$\xi_0$ stays independent of angle is to require that $\partial_t
\xi_0 = 0$.  We also require $\beta^A \doteq 0$, in order that the
angular coordinates be propagated along the null generators of
$\ScriPlus$.  These conditions greatly simplify the identification of
the Bondi-Sachs mass aspect from the asymptotic behavior, since they
make our coordinates close to {\em inertial} Bondi coordinates, for
which the Bondi-Sachs mass aspect can be read off directly from the
asymptotic expansion of the Bondi-Sachs metric.
 
The evolution equation for $\xi$, obtained from the evolution 
equation for the determinant of the angular part of the conformal 
metric in terms of the conformal extrinsic curvature, is
\begin{equation}
    \partial_t (\log{\xi}) = \beta^k \partial_k (\log{\xi}) -\frac{1}{2} 
    \beta^C{}_{\breve{\scriptscriptstyle{|}}C} + \tilde{\alpha} 
    \left( \frac{1}{2} \hat{\tilde{K}}^z_z - \frac{1}{3} \tilde{K} \right),
    \label{Eq:xievol}
\end{equation}
where $\beta^C{}_{\breve{\scriptscriptstyle{|}}C}$ is the divergence 
of the angular part of the shift considered as a vector field on the unit 
sphere.  At $\dot{\ScriPlus}$, this becomes 
\begin{equation}
    \partial_t(\log{\xi}) \doteq \tilde{\alpha}_0 \left( \frac{1}{2} 
    \tilde{\kappa} - \frac{1}{3} \tilde{K} \right).
\end{equation}
Therefore, we impose the boundary condition 
\begin{equation}
    \tilde{K} \doteq \frac{3}{2} \tilde{\kappa}_0.  
    \label{Eq:Ktildebc}
\end{equation}
We make no other restrictions on the conformal gauge, i.e., on the
details of the equation for $\tilde{K}$.  As long as the equation is
elliptic, as in Refs.~\cite{Bardeen2011a} and~\cite{Moncrief2009}, the
boundary condition is easily enforced.  If the conformal gauge is
evolved as part of a purely hyperbolic system, in which there are no
boundary conditions at $\ScriPlus$, things may not be so simple.

\section{The Bondi-Sachs energy and momentum}
\label{Sec:3}
The Bondi-Sachs energy is the true physical energy defined at future
null infinity in asymptotically flat spacetimes.  Its original
definition~\cite{Bondi1962,Sachs1962} was based on a special class of
retarded null coordinates $\left(u,r,\bar{\theta},\bar{\phi}\right)$
with a metric of the form, in close to the notation of
Ref.~\cite{Andersson1994},
\[
    ds^2 = - Ve^{2\beta } du^2 - 2e^{2\beta } dudr + r^2
    \bar{h}_{AB} \left( {d\bar{x}^{A} -
    U^{A} du} \right) \left( {d\bar{x}^{B} -
    U^{B} du} \right),
\]
where $\bar{h}_{AB}$ is a 2D metric asymptotically approaching the
unit sphere metric $\breve{h}_{AB}$ and whose determinant equals the
determinant of $\breve{h}_{AB}$ everywhere.  We put bars on the
angular coordinates to distinguish them from the angular coordinates
in the CMC hypersurface.  The expansion of $\bar{h}_{AB}$ in powers of
$x \equiv r^{-1}$ away from $\ScriPlus$  has the form
\begin{equation}
    \bar{h}_{AB} = \breve{h}_{AB} + 
    \bar{\chi}_{AB} x + \frac{1}{4} 
    \bar{\chi}^{CD}\,
    \bar{\chi}_{CD}\,\breve{h}_{AB} x^2
    + O(x^3).
    \label{Eq:hexp}
\end{equation}
The absence of any traceless contribution to the $O(x^2)$ term  in 
Eq.~(\ref{Eq:hexp}) implies the vanishing of the Weyl tensor on 
$\ScriPlus$, as Eq.~(\ref{Eq:zeroWeylCMC}) did on CMC hypersurfaces.
The remaining metric functions are asymptotically
\begin{equation}
    \label{Eq:V}
    V = 1 - 2\,M \left( u,\bar{\theta} ,\bar{\varphi} \right)\,x + O(x^2),
\end{equation}
\begin{equation}
    \label{Eq:beta}
    \beta = - \frac{1} {{32}}\,\bar{\chi}^{BC} \,
    \bar{\chi}_{BC}\, x^2 + O(x^3),
\end{equation}
\begin{equation}
    \label{Eq:U^A}
    U^{A} = - \frac{1}{2}\,\bar{\chi}^{AB}{}_
    {\breve{\scriptscriptstyle{|}}{B}} \, x^2 + O(x^3).
\end{equation}
Like $\breve{\chi}_{AB}$, $\bar{\chi}_{AB} \left( u,
  \bar{\theta},\bar{\varphi} \right)$ is a traceless symmetric tensor
on the unit sphere.  Its retarded time derivative along $\ScriPlus$ 
is the Bondi news.

The Bondi-Sachs metric function $M\left( u,\bar{\theta},\bar{\varphi}
\right)$ is what was originally identified as the Bondi mass aspect
\cite{Bondi1962}.  Its average over the unit sphere at a fixed
retarded time $u$ is the Bondi-Sachs energy $E_{\rm BS}$ and the
average over the unit sphere weighted by $N^i$, where
\begin{equation}
    N^i \equiv (\sin{\theta} \cos{\varphi}, \sin{\theta} \sin{\varphi}, 
    \cos{\theta}), 
\end{equation}
is the $i$th component of of the Bondi-Sachs momentum
three-vector in the asymptotic Minkowski frame.  However, there are
other less coordinate-specific definitions of the mass aspect which
give the same Bondi-Sachs energy and momentum when averaged.  Some are
based on the asymptotic behavior of Weyl tensor
\cite{Sachs1962,Penrose1964, TamburinoWinicour1966,Stewart1989} and
some are based on charge integrals derived from a Hamiltonian formalism
(see, e.g.,~\cite{Chrusciel1998,Chrusciel2004}).  What we will call
the Bondi-Sachs mass aspect is not the metric function $M$, but rather
the quantity which is monotonically decreasing on each of the null
generators of $\ScriPlus$~\cite{Chrusciel1998},
\begin{equation}
    M_{\rm A} \equiv M - \frac{1}{4} \bar{\chi}^{AB}{}_
    {\breve{\scriptscriptstyle{|}}AB}.
    \label{Eq:MAspectdef}
\end{equation}
The monopole and dipole angular moments of $M_{\rm A}$ and
$M$ are identical for arbitrary $\bar{\chi}^{\rm AB}$~\cite{Chrusciel2004}.

The task of actually calculating $M_{\rm A}$
from data on a CMC hypersurface can be accomplished in various ways.
The charge integrals of Ref.~\cite{Chrusciel2004} require comparing
the actual spatial metric and extrinsic curvature on the hypersurface
with a background metric and extrinsic curvature.  The results are
sensitive to the choice of background quantities.  The only fail-safe
procedure mentioned in Ref.~\cite{Chrusciel2004} is to find the
coordinate transformation from the CMC coordinates to Bondi
coordinates, where the appropriate background spacetime metric is
known, and then perform the inverse of this coordinate transformation
on the Bondi-Sachs background metric.  However, once one has
transformed to Bondi coordinates, it is much simpler just to read off
the Bondi-Sachs metric function $M$ from the transformed metric.
It is the latter procedure that we will follow.  While the coordinate
transformation involves the lapse and shift for the CMC coordinates,
the final result only depends on the spatial metric
and extrinsic curvature of a single CMC hypersurface.

The lapse and shift which preserve the CMC slicing condition and the
Gaussian normal coordinates based on $\dot{\ScriPlus}$ were derived in
Appendix A of Ref.~\cite{Bardeen2011a}.  The singular terms in the
elliptic equation for the conformal lapse allow no freedom in the
leading terms of the expansion away from $\dot{\ScriPlus}$, with the
result
\begin{equation}
  \tilde{\alpha} = \tilde{\alpha}_0\left[ 1 - \frac{1}{2}\tilde{\kappa}_0 z 
    + \frac{1}{4}\left( \frac{1}{2}\tilde{\kappa}_0^2 - \tilde{\kappa}_1 
      - 3\breve{\chi}^{C}_{\,D}
      \breve{\chi}^{D}_{\,C}
      + 2\xi_0^2 \right) z^2 + O \left( z^3 \right) \right].
\label{Eq:ConfLapseExpansion}  
\end{equation}
The coordinate components of the shift are, using the boundary
condition on $\tilde{K}$ of Eq.~(\ref{Eq:Ktildebc}),
\begin{equation}
  \beta^z = \tilde{\alpha}_0 \left( 1 -  \frac{1}{2} \tilde{\kappa}_0 
      z + O(z^2) \right)
     \label{Eq:betazshift}
\end{equation}
and
\begin{equation}
  \beta^{A} = \tilde{\alpha}_0 \xi_0^2 \left[ \left(
      \frac{1}{4} \breve{h}^{AB}\partial_{B}
      \tilde{\kappa}_0 - 
      \breve{\chi}^{AB}{}_{\breve{\scriptscriptstyle{|}}
        {B}} \right) z^2 + O(z^3) \right]. 
    \label{Eq:betaAshift}
\end{equation}

\subsection{Coordinate transformation to Bondi-Sachs coordinates}
\label{SubSec:3A}

The coordinate transformation to Bondi-Sachs coordinates is obtained
from the transformation equations for the inverse metric.  We
replace $r$ by its inverse $x$, so the Bondi-Sachs coordinates are
$\left( u,x,\bar{x}^A \right)$.  The inverse Bondi-Sachs metric
components, denoted by bars, are then
\begin{equation}
     \label{Eq:InverseBSMetric}
     \bar{g}^{uu} = \bar{g}^{u{A}} =0,
     \,\,\,\, \bar{g}^{ux} = x^2 e^{ - 2\beta }, \,\,\,\,
     \bar{g}^{xx} = x^4 Ve^{ - 2\beta },\,\,\,\, \bar{g}^{x{A}} =
     x^2 U^{A} e^{ - 2\beta }, \,\,\,\, \bar{g}^{AB} 
     = x^2 \bar{h}^{AB}.
\end{equation}
The metric transformation equations are in part 
partial differential equations for the Bondi-Sachs coordinates as functions 
of the CMC coordinates $\left( t,z,x^A \right)$ and in part equations for
the Bondi-Sachs metric functions.  We solve these equations term by
term in expansions in powers of $z$ away from $\dot{\ScriPlus}$ in the
CMC hypersurface.  From
\[
\bar{g}^{uu} = \frac{\Omega^2}{\tilde{\alpha}^2} \left[ - \left(
    \frac{\partial u}{\partial t} \right)^2 + 2 \frac{\partial
    u}{\partial t} \frac{\partial u}{\partial z} \beta^z + \left(
    \frac{\partial u}{\partial z} \right)^2 \left( \tilde{\alpha}^2 -
    \left( \beta^z \right)^2 \right) + O(z^4) \right] = 0,
\]
we get 
\[
    \frac{\partial u}{\partial t} = \left( \tilde{\alpha} + \beta^z \right) 
   \frac{\partial u}{\partial z} + O(z^4).
\]
The solution satisfying the boundary condition $u \doteq t$ is 
\begin{equation}
     u = t + \frac{z}{2\tilde{\alpha}_0} \left[ 1 + \frac{1}{4} 
     \tilde{\kappa}_0 z + O(z^2) \right].
     \label{Eq:utrans}
\end{equation}
From 
\[
\bar{g}^{uA} = \frac{\Omega^2}{\tilde{\alpha}^2} \left[ \beta^z
  \frac{\partial \bar{x}^A}{\partial z} - \beta^A \beta^z
  \frac{\partial u}{\partial z} + \tilde{h}^{AC} \frac{\partial u}
  {\partial x^C} + O(z^3) \right] = 0
\]
and 
\[
    \frac{\partial u}{\partial x^C} = \frac{1}{8\tilde{\alpha}_0} 
    \frac{\partial \tilde{\kappa}_0}{\partial x^C} z^2  + O(z^3),
\]
with the boundary condition $\bar{x}^A \doteq x^A$,
\begin{equation}
    \bar{x}^A = x^A + \frac{1}{6}  \xi_0{}^2 \left( 
    \breve{\chi}^{AB}{}_{\breve{\scriptscriptstyle{|}} B} - 
    \frac{1}{2} \tilde{\kappa}_0{}^{\breve{\scriptscriptstyle{|}} A} 
    \right) z^3 + O(z^4).
    \label{Eq:xAtrans}
\end{equation}
The Bondi-Sachs $x$-coordinate is obtained from the transformation of
the angular part of the inverse metric,
\begin{equation}
    x^2 \bar{h}^{AB} = \Omega^2 \left[ 
    \tilde{h}^{CD} \frac{\partial \bar{x}^A}{\partial x^C} 
    \frac{\partial \bar{x}^B}{\partial x^D} + O(z^4) \right].
    \label{Eq:hABtrans}
\end{equation}
Equating the determinants of the two sides gives 
\begin{equation}
    x = \Omega \xi \left[ 1 +  \frac{1}{12} \xi_0{}^2 \left( 
    \breve{\chi}^{AB}{}_{\breve{\scriptscriptstyle{|}} AB} - 
    \frac{1}{2} \breve{\Delta} \tilde{\kappa}_0 \right) z^3 + 
    O(z^4) \right],
    \label{Eq:xvz}
\end{equation}
where $\breve{\Delta}$ is the Laplacian operator on the unit sphere.
Substituting back into Eq.~(\ref{Eq:hABtrans}) gives
\begin{equation}
    \bar{\chi}^{AB} = - 2 \left( \frac{z}{\Omega \xi} \right)_0 
    \breve{\chi}^{AB} = - 2 \left (\frac{3}{K \xi_0} \right) \breve{\chi}^{AB}.
    \label{Eq:chiABtrans}
\end{equation}
The second-order contribution to $\tilde{h}^{AB}$ is consistent with
the second-order contribution to $\bar{h}^{AB}$ once the correction is
made for the time dependence of $\breve{\chi}^{AB}$ (Eq.~(A34) of
Ref.~\cite{Bardeen2011a}), taking into account the difference between
$t$ and $u$ at the same physical point in the interior of the
hypersurface (Eq.~(\ref{Eq:utrans})).

The Bondi-Sachs metric functions $\beta$ and $U^A$ obtained from the
coordinate transformation are consistent with the Bondi-Sachs
asymptotic solutions from the Einstein equations, taking into account
Eq.~(\ref{Eq:chiABtrans}).  To obtain the metric function $V$, from
which the mass aspect is derived, we use Eq.~(\ref{Eq:xvz}) and the
fact that $\tilde{\alpha} - \beta^z = O(z^2)$ to get
\begin{equation}
  x^4 V e^{-2\beta} = \left( \frac{\Omega}{\tilde{\alpha}} \right)^2 
  \left\{ 2\beta^z \partial_z \left(\Omega \xi \right) \left[ \partial_t 
      \left( \Omega \xi \right)+ \left( \tilde{\alpha} - \beta^z \right) 
    \partial_z \left( \Omega \xi \right) \right] + O(z^4) \right\}.
    \label{Eq:gbarxx1}
\end{equation}
From Eqs.~(\ref{Eq:Omegaevol},~\ref{Eq:xievol}), the time derivative
of $\Omega \xi$ is
\begin{equation}
    \partial_t \left(\Omega \xi \right) = \beta^z \partial_z 
    \left(\Omega \xi \right) - \frac{K}{3} \tilde{\alpha} \xi + \beta^A 
    \partial_A \left(\Omega \xi \right) + \frac{1}{2} \Omega \xi \left( 
    \tilde{\alpha} \hat{\tilde{K}}_z^z - 
    \breve{\nabla}_c \beta^c \right).
    \label{Eq:Omxit1}
\end{equation}
Note that $\tilde{K}$ cancels out, so there is no direct dependence on 
the conformal gauge.  Inserting Eq.~(\ref{Eq:Omxit1}) into 
Eq.~(\ref{Eq:gbarxx1}) and rearranging terms gives 
\begin{equation}
  x^4 V e^{-2\beta} = \left( \frac{\Omega^2}{\tilde{\alpha}} \right)^2 
  2\beta^z \partial_z \left( \Omega \xi \right) \left[ \beta^z \left( 
    \partial_z \left( \Omega \xi \right) - \frac{K}{3} \xi \right) + 
    \Omega \xi \left( \tilde{\alpha} \hat{\tilde{K}}^z_z - 
    \breve{\nabla}_c \beta^c \right) + O(z^4) \right].  
    \label{Eq:gbarxx2}                                                                                                                                                                                                                          
\end{equation}
From the expansions of $\Omega$ and $\xi$ in
Eqs.~(\ref{Eq:AsympOmega}, \ref{Eq:xiexp}),
\begin{eqnarray}
  \Omega \xi &=& \frac{K \xi_0}{3} z \left[  1 + \frac{1}{4} \tilde{\kappa}_0 z 
    + \frac{1}{6} \left( \xi_0{}^2 + \frac{1}{4} \tilde{\kappa}_0{}^2 + 
      \frac{1}{2} \tilde{\kappa}_1 - \breve{\chi}^A_B \breve{\chi}^B_A \right) 
    z^2 \right. \nonumber \\
  &+& \left. \left( c_3 + \frac{1}{8} Q \log{\frac{K z}{3}} + \frac{1}{12} 
      \xi_0{}^2 \tilde{\kappa}_0 + \frac{1}{96} \tilde{\kappa}_0{}^3 
      - \frac{1}{48} \tilde{\kappa}_0 \tilde{\kappa}_1 + \frac{1}{6} 
      \tilde{\kappa}_2 - \frac{1}{12} \tilde{\kappa}_0 \breve{\chi}^A_B 
      \breve{\chi}^B_A \right) z^3 + O(z^4) \right].
    \label{Eq:Omxiexp}
\end{eqnarray}
Combine Eqs.~(\ref{Eq:gbarxx2}), (\ref{Eq:Omxiexp}), (\ref{Eq:xiexp}),
(\ref{Eq:AsympKzz}), (\ref{Eq:betaAshift}), as well as the first-order
expansions of $\Omega$, $\tilde{\alpha}$, and $\beta^z$, and simplify
to get
\begin{equation}
  V = 1 + \frac{3}{K \xi_0{}^3} \left[ 8 c_3 + d_1 + \frac{1}{4} Q - 
    \frac{1}{3} \xi_0{}^2 \tilde{\kappa}_0 + \frac{1}{24} \tilde{\kappa}_0{}^3 - 
    \frac{5}{12} \tilde{\kappa}_0 \left( \tilde{\kappa}_1 + 
   \breve{\chi}^C_D \breve{\chi}^D_C \right) +  \tilde{\kappa}_2 + 
    \xi_0{}^2 \left( \breve{\chi}^{CD}{}_{\breve{\scriptscriptstyle{|}}CD} - 
      \frac{1}{4} \breve{\Delta} \tilde{\kappa}_0 \right) \right] 
  \Omega \xi + O(z^2).
    \label{Eq:Vexp}
\end{equation}
The Bondi-Sachs mass aspect $M_{\rm A}$ as defined in
Eq.~(\ref{Eq:MAspectdef}) is
\begin{equation}
  M_{\rm A} = - \frac{3}{K \xi_0{}^3} \left[ 4 c_3 + \frac{1}{2} d_1 + 
    \frac{1}{8} Q - \frac{1}{6} \xi_0{}^2 \tilde{\kappa}_0 + \frac{1}{48} 
    \tilde{\kappa}_0{}^3 - \frac{5}{24} \tilde{\kappa}_0 \left( \tilde{\kappa}_1 + 
    \breve{\chi}^C_D \breve{\chi}^D_C \right) + \frac{1}{2} \tilde{\kappa}_2 
     - \frac{1}{8} \xi_0{}^2 \breve{\Delta} \tilde{\kappa}_0 \right],
    \label{Eq:MassAspect}
\end{equation}
with $Q$ defined by Eq.~(\ref{Eq:Qdef}), which is equivalent to
\begin{equation}
    Q = \xi_0{}^2 \breve{\chi}^{CD}{}_{\breve{\scriptscriptstyle{|}}CD} + 
    \breve{\chi}^C_D \partial_t \breve{\chi}^D_C.
    \label{Eq:Qdefb}
\end{equation}

The alternative approach to calculating the Bondi-Sachs mass aspect
based on the Weyl scalar $\Psi_2$ in the Newman-Penrose formalism is
also straightforward to implement.  The real part of $\Psi_2$ is the
component $\hat{\tilde{E}}^z_z = s^k s^l \hat{\tilde{E}}_{ kl}$, where
$s^k$ is the unit outward normal to the $\dot{\ScriPlus}$ 2-surface,
of the traceless electric part of the conformal Weyl tensor.  Its
normal derivative at $\dot{\ScriPlus}$ is related to the mass aspect
by
\begin{equation}
    \frac{K \xi_0{}^3}{3} M_{\rm A} - \frac{1}{2} \xi_0{}^2 
    \breve{\chi}^{CD}{}_{\breve{\scriptscriptstyle{|}}CD} = 
    \frac{K \xi_0{}^3}{3} M = -\frac{1}{2} \left[ \partial_z 
    \hat{\tilde{E}}^z_z + \breve{\chi}^C_D \partial_t \breve{\chi}^D_C 
    \right]_{\dot{\ScriPlus}}.
    \label{Eq:Weylformula}
\end{equation}
This simple form is valid if the Weyl tensor vanishes on $\ScriPlus$ 
and the gauge evolution boundary conditions preserve the 
explicit spherical geometry of $\dot{\ScriPlus}$, with angular 
coordinates propagated along the null generators, such that the 
curves on $\ScriPlus$ with tangent vector $\partial_t$ are null 
geodesics.  A 3D expression for $\hat{\tilde{E}}_{ij}$, from 
Appendix A of Ref.~\cite{Bardeen2011a}, is 
\begin{equation}
    \hat{\tilde{E}}_{ij} = \frac{1}{\Omega} \left \{ \tilde{\nabla}_i 
    \tilde{\nabla}_j \Omega + \frac{K}{3} \hat{\tilde{K}}_{ij} \right \}^{TF} 
    + \left \{ \hat{\tilde{R}}_{ij} - \hat{\tilde{K}}_i^k \hat{\tilde{K}}_{kj} 
    \right \}^{TF} .
    \label{Eq:WeylE}
\end{equation}

\subsection{Conformally flat initial data}
\label{SubSec:3B}

The general expression derived in the previous section simplifies
drastically for conformally flat initial data.  By conformal flatness we
mean that the conformal spatial metric is Euclidean, $\tilde{g}_{ij} =
\delta_{ij}$, but we still assume a CMC hypersurface.  Future null
infinity is a coordinate sphere,
\[
R \equiv \sqrt{ (x^1)^2 + (x^2)^2 + (x^3)^2} = R_+.
\]
The Gaussian normal coordinates based on $\dot{\ScriPlus}$ are $z =
R_+ - R$ and the usual polar angles $x^A$ of spherical coordinates.
The extrinsic curvature of constant-$R$ two-surfaces is isotropic,
with $\tilde{\kappa} = 2/R = 2/(R_+ - z)$, and
\[
\tilde{\kappa}_0 = \frac{2}{R_+} = 2 \xi_0, \qquad \tilde{\kappa}_1 =
2 \xi_0{}^2, \qquad \tilde{\kappa}_2 = 2 \xi_0{}^3.
\]
No gravitational waves are present at $\dot{\ScriPlus}$, since
$\breve{\chi}_{AB} = \breve{\psi}_{AB} = 0$.

For the conformal extrinsic curvature on the CMC hypersurface, we
adopt the generalized Bowen-York solutions of the vacuum momentum
constraint equations from Ref.~\cite{Buchman:2009ew}:
\begin{eqnarray}
    \label{Eq:GeneralizedBY}
    \Omega^{-2} \hat{\tilde{K}}_{ij} \equiv \tilde A_{ij} &=& 
    \frac{C} {{R_{\mathbf{D}}^3 }}\left[ {3n_i n_j -\delta _{ij} } \right] - \frac{3}
    {{R_{\mathbf{D}}^3 }}\left[ {\varepsilon _{ik{\ell}} S^k n^{\ell} n_j
    + \varepsilon _{jk{\ell}} S^k n^{\ell} n_i } \right] \nonumber \\
    &&- \frac{3}{{2R_{\mathbf{D}} ^2 }}
    \left[ {P_i n_j+P_j n_i+P^kn_k\left( {n_i n_j-\delta _{ij}} \right)} \right] \nonumber \\
    &&+ \frac{3}{{2R_{\mathbf{D}} ^4 }}
    \left[ {Q_i n_j  + Q_j n_i  + Q^k n_k \left( {\delta _{ij}  - 5n_i n_j } \right)} \right], 
\end{eqnarray} 
with $R_{\mathbf{D}} \equiv \left| {{\mathbf{x}}- {\mathbf{D}}}
\right|$, $n^i \equiv \left( {x^i - D^i }\right)/R_{\mathbf{D}} $.
Since the momentum constraint equations are linear, these solutions
can be superimposed to generate initial data for multiple individual
black holes or a single distorted black hole, depending on whether,
after solving for the conformal factor, there are disjoint apparent
horizons or a single all-encompassing apparent horizon.  We follow
Ref.~\cite{Buchman:2009ew} in only considering cases for which each
solution represents a separate black hole centered at the coordinate
position $x^i = D^i$, and impose an inner excision boundary condition
for each black hole so that the coordinate sphere $R_{\mathbf{D}} =
R_{\rm ms}$ is a minimal 2-surface.  The minimal surface condition,
that the normal derivative of $\Omega$ vanish, is an inner boundary
condition on the elliptic equation for the conformal factor.  The
choice of $R_{\rm ms}$ is an additional input parameter for each black
hole.  In the limit $R_{\rm ms} \to 0$, the Einstein-Rosen bridge
associated with the minimal surface becomes an infinitely long
``trumpet'' configuration.  The minimal surface should be inside the
apparent horizon, which means $R_{\rm ms}$ should not be too large and
also requires $C > 0$.

For black holes on a conventional asymptotically flat (e.g. maximal)
hypersurface, with the boundary condition $\Omega \to 1$ as $R \to
\infty$, the sum of the boost 3-vectors $\mathbf P$ is the ADM linear
momentum of the system.  The physical interpretation of the boost is
not so straightforward on CMC hypersurfaces.  The vector $\mathbf Q$
represents the boost of the black hole as viewed from the other side
of its Einstein-Rosen bridge.  For a single centered black hole, it
can be related to $\mathbf P$ through a condition of inversion
symmetry about the minimal surface, but initial data without inversion
symmetry are also valid.  In our numerical results we will take
$\mathbf Q = 0$.  Note that $\tilde A_{ij} $ is finite and non-zero at
future null infinity, $R = R_ +$, which means that the traceless part
of the physical extrinsic curvature goes to zero as $\Omega^3$ 
for all contributions.  This is quite unlike the situation on maximal
hypersurfaces.

Since conformal flatness implies that the conformal Ricci tensor
vanishes, the equation for $\Omega$ simplifies to
\begin{equation}
    \label{Eq:HamConstCF}
    \Omega \tilde{\Delta} \Omega = \frac{3}{2} \left( \tilde \nabla _k 
    \Omega\,\tilde \nabla ^k \Omega - \frac{K^2}{9} \right)
    +  \frac{{\Omega ^6 }} {4}\tilde A_{ij} \tilde A^{ij},
\end{equation}
with $\tilde{\Delta}$ the flat space Laplacian operator.  The equation
is elliptic almost everywhere, but is degenerate at the outer boundary
$R = R_+$, where the boundary condition $\Omega = 0$ is imposed.
Solutions are characterized by the values of $K$ and $R_+$ and, for
each black hole, the Bowen-York parameters and $R_{\rm ms}$.

A natural dimensionless parameter for expanding the solution away from
$\dot{\ScriPlus}$ is $\bar z \equiv \xi_0 z$.  The asymptotic solution
of Eq.~(\ref{Eq:HamConstCF}) then reduces to
\begin{equation}
    \Omega = \frac{K R_+}{3} \bar{z} \left[ 1 - \frac{1}{2} \bar{z} + 
    \bar{c}_3 \bar{z}^3 \left(  1 + \bar{z} \right) + 
    O(\bar{z}^5) \right]. 
    \label{Eq:AsympOmCF}
\end{equation}
Here $\bar{c}_3 \equiv R_{+}^3 c_3$ is the rescaled dimensionless
version of the locally undetermined coefficient $c_3$ of
Eq.~(\ref{Eq:AsympOmega}). The angular derivatives in $\tilde{\Delta}
\Omega$ first affect the solution in $O(\bar{z}^5)$ and the conformal
extrinsic curvature first contributes in $O(\bar{z}^6)$.

With conformal flatness plus the regularity conditions, the 3D
conformal extrinsic curvature has
\[
    \hat{\tilde{K}}^z_z = \Omega^2 N^i \tilde{A}_{ij} N^j = d_1 z^2,
\]
and we define the rescaled dimensionless form of $d_1$ as 
\begin{equation}
    \bar{d}_1 \equiv R_+^3 d_1 \equiv 2 \bar{A}_{RR} = 
    \left(\frac{K}{3} \right)^2 R_+^3 N^i \tilde{A}_{ij} N^j.
    \label{Eq:ARRdef}
\end{equation}
Eq.~(\ref{Eq:MassAspect}) for the Bondi-Sachs mass aspect simplifies to 
\begin{equation}
  M_{\rm A} = - \frac{3}{K}\left( 4 \bar{c}_3 + \frac{1}{2} \bar{d}_1
  \right) = - \frac{3}{K}\left( 4 \bar{c}_3 + \bar{A}_{RR} 
  \right).
  \label{Eq:BSMassAspect}
\end{equation}
Note that for spherical symmetry in Eq.~(\ref{Eq:GeneralizedBY}), all
the Bowen-York parameters except $C$ are zero and $\bar{A}_{RR} =
\left(K/3 \right)^2 C \equiv \bar{C}$.  See Ref.~\cite{Buchman:2009ew}
for an extensive discussion of CMC hypersurfaces in the Schwarzschild
geometry and the significance of the parameter $\bar{C}$.  The two
contributions to the Bondi-Sachs mass aspect are labeled by $\left(
  M_{\rm A} \right)_{\rm \Omega}$ for the contribution from the
conformal factor and by $\left( M_{\rm A} \right)_{\rm K}$ for the
contribution from the conformal extrinsic curvature.

To sum up, the Bondi-Sachs energy and 3-momentum associated with a CMC
hypersurface are given by the integrals on $\dot{\ScriPlus}$: 
\begin{equation}
  \label{Eq:BondiEnergy}
  E_{\rm BS}  = \frac{1}{{4\pi }}
  \oint {M_{\rm A} \sqrt {\det \breve{h}_{AB} } } \,\,dx^{A} dx^{B} 
  =\frac{1}{{4\pi }}\oint {M_{\rm A} } \sin \theta d\theta d\phi 
\end{equation}
and 
\begin{equation}
    \label{Eq:3-Momentum}
    P_{\rm BS}^k  = \frac{1}{{4\pi }}
    \oint {M_{\rm A} } N^k \sin \theta d\theta d\phi .
\end{equation}
An expression for the angular momentum as an integral over
$\dot{\ScriPlus}$ in the context of Bondi coordinates is given in
Sec.~6 of Ref.~\cite{Chrusciel2004}.  In the absence of radiation, and
specifically in the conformally flat case with Euclidean coordinates
on a CMC hypersurface, this becomes
\begin{equation}
  \label{Eq:AngMom}
  J_i = - \frac{1} {{8\pi }}\oint\limits_{R = R_ + } {\sin \theta \,
    d\theta \, d\phi \,\varepsilon _{ijk} \,N^j \left( {R^3 \tilde
        A_\ell^{~k} } \right)N^\ell } .
\end{equation}

Analytic expressions for the $\left( M_{\rm A} \right)_{\rm K}$ 
contributions to the Bondi-Sachs energy and linear momentum in terms
of the Bowen-York parameters of Eq.~(\ref{Eq:GeneralizedBY}) are
obtained in Appendix~\ref{AppA}, and the angular momentum integral is
evaluated in Appendix~\ref{AppB}.  The result for the angular momentum
3-vector $\mathbf J$ is deceptively Newtonian,
\begin{equation}
     \label{Eq:AnalAngMom}
    \mathbf J = \mathbf S + \mathbf D \times \mathbf P,
\end{equation}
deceptive because $\mathbf D$ is a coordinate displacement, not a physical 
distance, and the boost $\mathbf P$ is not a physical linear momentum.

\section{Numerical evaluation of the Bondi-Sachs 
energy and momentum}
\label{Sec:4}

For conformally flat initial data, with an analytic solution of the
conformal momentum constraint equation such as that of
Eq.~(\ref{Eq:GeneralizedBY}), the only significant numerical task is
solving the Hamiltonian constraint equation for the conformal factor
$\Omega$.  Our results are obtained with the pseudospectral elliptic
solver that is part of the Caltech-Cornell-CITA Spectral Einstein Code
({\tt SpEC}), described in detail in
Refs.~\cite{SpECwebsite,Pfeiffer2003}. The overall domain is divided
into a number of sub-domains, which are spherical near the boundary at
$R=R_+$.  The output for the conformal factor in the outermost
spherical sub-domain, from an inner radius $R_1$ to $R_+$, is in the
form of coefficients of a spherical harmonic expansion $\Omega =
\sum_{\ell,m} \Omega_{\ell m} Y_\ell^m$ at each of a number of radial
collocation points.

To obtain the coefficients of the spherical harmonic expansion of
$\bar{c}_3$, we fit the asymptotic form of Eq.~(\ref{Eq:AsympOmCF}) to
the numerical results for the conformal factor $\Omega$.  It is
counterproductive to use collocation points too close to $R_+$ where
the small contribution of the $\bar{c}_3$ terms to $\Omega$ may not be
large compared to numerical error, so we exclude collocation points
with $\bar{z} < \bar{z}_2 $ and fit
\[
    \frac{1}{\left(1 + \bar{z} \right) \bar{z}^4} \left( \frac{3}{K R_+}
    \Omega_{\ell m} - \bar{z} + \frac{1}{2} \bar{z}^2 \right)
\]
to the form
\[
    \left( \bar{c}_3 \right)_{\ell m}  + e_{\ell m} \,\bar{z}^2 + 
    f_{\ell m} \,\bar{z}^3 
\]
using a standard
Grace\footnote{http://plasma-gate.weizmann.ac.il/Grace/} curve-fitting
routine.  In practice, for single black holes, we find that taking
$R_1/R_ + = 0.95$ and $\bar{z}_2 \gtrsim 0.02$ gives values of $\left(
  \bar{c}_3 \right)_{00}$ stable to about $1$ part in $10^5$, as long
as the black hole centers are within $0.12\,R_+$ or so, and the boost
parameters are not too large.  We also need the coefficients of the
spherical harmonic expansion of $\bar{A}_{RR}$, which are obtained
analytically in Appendix~\ref{AppA} for $\ell = 0$ and $\ell = 1$.
Then the Bondi-Sachs energy is
\begin{equation}
\label{Eq:BondiEnergyNum}
E_{\rm BS} = - \frac{3}{K} \left[4\left( \bar{c}_3 \right)_{00} +
  \left( \bar{A}_{RR} \right)_{00} \right] \sqrt{\frac{1}{4\pi}}.
\end{equation}

The Cartesian components of the radial unit vector $N^i$ are composed
only of $\ell = 1$ spherical harmonics, and since $\bar{c}_3$ is real,
the standard Condon-Shortley phase convention gives $\left( \bar{c}_3
\right)_{1\,-1} = - \left( \bar{c}_3 \right)_{1\,1}$.  Then from
Eq.~(\ref{Eq:3-Momentum}) for the Cartesian components of the
Bondi-Sachs momentum,
\begin{eqnarray}
  \label{Eq:3-MomentumNum}
  P^x_{\rm{BS}} &=& + \frac{3} {K}\operatorname{Re} \left[ {4\left(
        {\bar{c}_3 } \right)_{11} + \left( \bar{A}_{RR} \right)_{11} } 
  \right] \sqrt{\frac{1}{6 \pi } }, \nonumber \\
  P^y_{\rm{BS}} &=& - \frac{3} {K}\operatorname{Im} \left[ {4\left( 
        {\bar{c}_3 } \right)_{11} + \left( \bar{A}_{RR} \right)_{11} } 
  \right] \sqrt{\frac{1}{6 \pi }} , \\
  P^z_{\rm{BS}} &=& - \frac{3} {K} \left[ {4\left( \bar{c}_3 \right)_{10} 
      + \left( \bar{A}_{RR} \right)_{10} } \right] \sqrt{\frac{1}
    {12 \pi}} . \nonumber 
\end{eqnarray} 

Analytic expressions for the contributions to the Bondi-Sachs energy
and $3$-momentum from the conformal extrinsic curvature, the
$\bar{A}_{RR}$ terms in Eqs. (\ref{Eq:BondiEnergyNum}) and
(\ref{Eq:3-MomentumNum}), and their dependence on the Bowen-York
parameters, are derived in Appendix~\ref{AppA}.  The results are given
in Eqs.~(\ref{Eq:EnergyBS-A}) and~(\ref{Eq:PvecBS-A}).  The total
Bondi-Sachs energy and $3$-momentum form a $4$-vector in the
asymptotic Minkowski spacetime, and the Bondi-Sachs mass of the system
is defined to be the special relativistic magnitude of this
$4$-vector,
\begin{equation}
\label{Eq:M_BS}
M_{\rm{BS}}=\sqrt {E_{\rm{BS}}^2 - \mathbf P_{\rm{BS}}^2 }.
\end{equation}
The angular momentum of the system has no contribution from the 
conformal factor and is given analytically by Eq.~(\ref{Eq:AnalAngMom}) 
(see Appendix~\ref{AppB}). 

There are two arbitrary scale factors in our problem, the choice of the 
unit of mass-length-time and in the definition of the conformal
factor.  Only combinations of variables invariant under both a change 
of units and a uniform conformal rescaling are physically 
meaningful.  Combinations of the input parameters with this property are:
\begin{equation}
  \bar{C} \equiv (K/3)^2 C, ~~ \bar{P}^i \equiv (K^2 R_+ /18) P^i, ~~
  \bar{Q}^i \equiv (K^2/18 R_+) Q^i, ~~ \bar{S}^i \equiv S^i/C, ~~ 
  \bar{R}_{\rm ms} \equiv R_{\rm ms}/R_+, ~~ \bar{D}^i \equiv D^i/R_+.  
\end{equation}
For sufficiently small $\bar{R}_{\rm ms}$, so there is a trumpet-like
configuration inside the apparent horizon, $\bar{C}^{\frac{1}{2}}
\approx (K/3) M$ for a Schwarzschild-like black hole of mass $M$ (see
Fig. 4 in Ref.~\cite{Buchman:2009ew}).  Making $\bar{C} \ll 1$ ensures
that the warping effects of the mean extrinsic curvature are small in
the vicinity of the apparent horizon of a centered black hole.  As
argued in Ref.~\cite{Buchman:2009ew}, when this is the case, the
physical momentum should be roughly $\Omega_{\rm max}$ times the
boost.  From the asymptotic solution for the conformal factor,
$\Omega_{\rm max} \approx \left| \partial \Omega / \partial R
\right|_{R_+} (R_+/2) = K R_+/6$, so that we expect $K R_+/6$ times
the Bowen-York boost (or $3/K$ times $\bar{P}^i$) to be a rough
estimate of the physical Bondi-Sachs 3-momentum of the black hole.
Inversion symmetry about the apparent horizon, as sometimes assumed,
implies $Q^i = \pm R_{\rm ms}^2 P^i$, so at $\ScriPlus$ the $Q^i$
terms in the conformal extrinsic curvature are of order $\bar{R}_{\rm
  ms}^2$ times the $P^i$ terms and typically make a negligible
contribution to the Bondi-Sachs momentum.  We will simply take $Q^i =
0$ in our numerical calculations.  Well away from any black hole
apparent horizons and not too close to $R_+$ ($R < R_+/3$), the
conformal factor plateaus near its maximum value, so in this regime
physical distances are roughly coordinate distances divided by $K
R_+/6$.

A key output parameter is the dimensionless rescaled ``irreducible
mass'' $\bar{M}_{\rm irr} \equiv (K/3) M_{\rm irr} = (K/3)
\sqrt{A/16\pi}$, where $A$ is the area of the black hole apparent
horizon.  The irreducible mass is used to scale the results for the
Bondi-Sachs energy, momentum, and mass.  In
Ref.~\cite{Buchman:2009ew}, the irreducible mass is used as a
surrogate for the physical mass in the discussion of results, so 
giving our results in units of the irreducible mass
facilitates comparison with that paper.

For single black holes, we vary the Bowen-York parameters starting
from a representative spherically symmetric model that has $\bar{C} =
0.0011207$, $\bar{R}_{\rm ms} = 0.00127$, and $\bar{M}_{\rm irr} =
0.028339$.  For these parameters, the effects of the non-zero $K$
start becoming important at about $10$ times the radius of the
apparent horizon, which means initial data near the black hole
are similar to initial data on a conventional maximal hypersurface.
We also present results for two examples of binary black hole initial
data.

\subsection{Single spinning black hole}
\label{SubSec:SingleSpinningBh}
\begin{figure}
 \includegraphics[width=0.4\columnwidth]{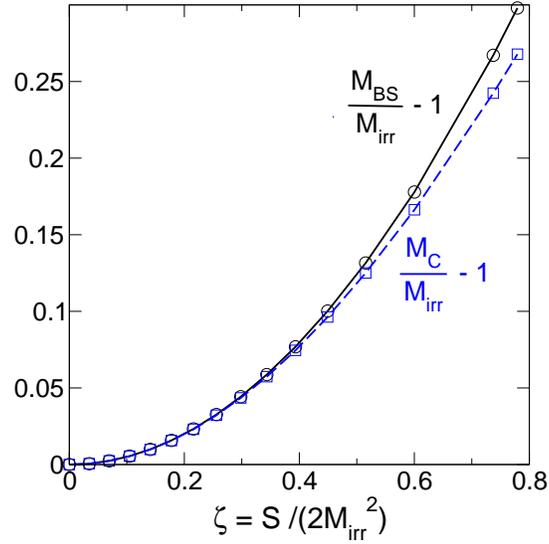}
 \caption{\label{fig:Diff_BE_Mc} Single spinning unboosted centered
   black hole: The ratio of the Bondi-Sachs mass ($M_{\rm BS}$) to the
   irreducible mass ($\Mirr$) is compared to the ratio of the
   Christodoulou mass ($M_{\rm C}$) to the irreducible mass, as the
   spin extremality parameter $\zeta$ is varied from $0$ to $0.78$. }
\end{figure}

Here we see how the Bondi-Sachs mass for a single centered black hole
with no Bowen-York boost varies with Bowen-York spin
$\mathbf{S} = (0,0,S)$.  As the dimensionless magnitude of
the spin $\bar S$ increases, we keep $\bar{C}_{\rm eff} \equiv \bar{C}
\sqrt{1 + 2\left(\bar{S}\right)^2}$ rather than $\bar{C}$
constant, because the square root of $\bar{C}_{\rm
  eff}$ tracks $\bar{M}_{\rm irr}$ much better than
$\bar{C}^{\frac{1}{2}}$ as $\bar S$ becomes larger than $1$.  There is
no linear momentum, so the Bondi-Sachs mass equals the Bondi-Sachs
energy, and by Eq.~(\ref{Eq:AnalAngMom}) the angular momentum equals
the spin.  The ratio $M_{\rm BS} / \Mirr -1$ is plotted in
Fig.~\ref{fig:Diff_BE_Mc} as a function of the spin extremality
parameter $\zeta \equiv S / (2 \Mirr^2)$, for $0 \le \zeta \le 0.78$.  
The sequence terminates at $\zeta = 0.78$ in the limit $\bar{S}
\to \infty$ for conformally flat initial data (see
Ref.~\cite{Buchman:2009ew}).  We also plot the corresponding quantity
for the Christodoulou mass~\cite{Christodoulou70} $M_{\rm C}$, which
is defined as the mass of the Kerr black hole with the same $\Mirr$
and angular momentum.  The Bondi-Sachs mass for our initial data is a
bit larger than the mass of the Kerr black hole for the same $\zeta$,
but the maximum $M_{\rm C} / \Mirr - 1$ is significantly larger,
$0.414$, because for a Kerr black hole the maximum $\zeta$ and the
maximum $J/M_{\rm C}^2$ are both $1$.

\subsection{Single displaced black hole with zero spin and boost}
\label{SubSec:SingleDisplacedBh}
\begin{figure}
 \includegraphics[width=0.4\columnwidth]{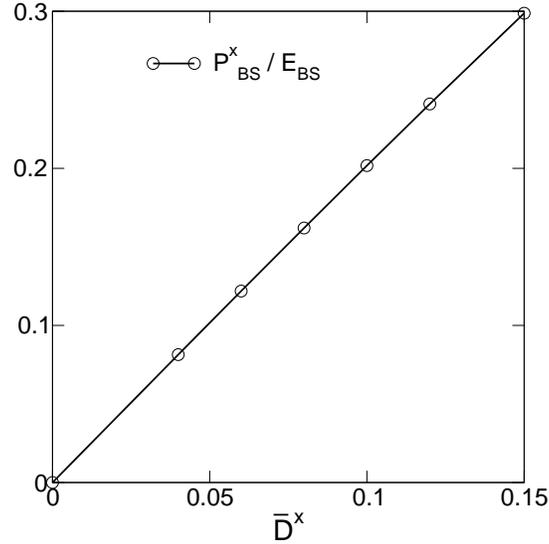}
 \caption{\label{fig:CxVsPx} Single black hole displaced in the
   $x$-direction with no boost or spin.  The velocity of the black
   hole in the asymptotic Minkowski frame, $|P^x_{\rm BS}|/E_{\rm
     BS}$, is plotted against the relative displacement $\bar{D}^x =
   D^x/R_+$.  The remaining input parameters ($\bar C$, $\bar{R}_{\rm
     ms}$) have their usual values. }
\end{figure}

Now we displace a single, non-spinning, {\em unboosted} black hole a
coordinate distance $D$ (in the $x$-direction) from the center of the
$\dot{\ScriPlus}$ coordinate sphere.  The other Bowen-York parameters
are the same as for our representative spherically symmetric black
hole.  Surprisingly, rather than being zero, the velocity of the black
hole in the asymptotic Minkowski frame (given by the ratio of the
Bondi-Sachs momentum to the Bondi-Sachs energy) is proportional to and
in the same direction as the displacement from the origin, as shown in
Fig.~\ref{fig:CxVsPx}. The dominant contributions to the Bondi-Sachs
energy and momentum are from the conformal factor, and have opposite
signs from the contributions associated directly with the extrinsic
curvature.  The contributions from the extrinsic curvature, as
calculated using Eqs.~(\ref{Eq:EnergyBS-A}) and~(\ref{Eq:PvecBS-A}),
are simply $(E_{\rm BS})_{\rm K} = - KC / 3$ and $(\mathbf{P}_{\rm
  BS})_{\rm K} = - (\mathbf{D}/R_+)KC/3$.  With $\bar{D} = 0.1$,
$\bar{M}_{\rm irr} = 0.028340$, and $\bar{C} = 0.0011207$, the
fractional extrinsic curvature contributions to the Bondi-Sachs energy
and momentum are, respectively, $(E_{\rm BS})_{\rm K}/E_{\rm BS} =
-0.039$ and $(\mathbf{P}_{\rm BS})_{\rm K}/\mathbf{P}_{\rm BS} =
-0.019$.

As an indication of whether the displaced black hole is just a moving
Schwarzschild black hole or whether it has additional distortions, we
compare the Bondi-Sachs mass as given by Eq.~(\ref{Eq:M_BS}) with the
irreducible mass derived from the area of the apparent horizon. We
find that these are the same to the numerical accuracy of our
calculations (about $5$ significant figures), which suggests that the
displaced black hole is in fact just a Schwarzschild black hole in an
asymptotic Lorentz frame in which the black hole is not at rest.  We
have not tried to prove this more decisively, by finding the
coordinate transformation to the conventional form of the
Schwarzschild metric.

\subsection{Single centered boosted black hole}
\label{SubSec:SingleCenteredBoostedBH}
Now consider a Bowen-York boost as the only deviation from
Schwarzschild.  On an asymptotically flat hypersurface, such as a
maximal hypersurface, the Bowen-York boost is equal to the physical
ADM momentum.  In the limit of small $K^2 C$, there is a range of $R$
over which $\Omega$ is roughly constant near its maximum and $M_{\rm
  irr} << R/\Omega << 1/K$, a region of the CMC hypersurface
approximating the asymptotic part of a maximal hypersurface.
Ref.~\cite{Buchman:2009ew} argued on this basis for using $\mathbf{P}
\,\Omega_{\rm max}$ as an estimate of the Bondi-Sachs linear
momentum. The close agreement between $\bar{P}^z$ and $(K/3)P_{\rm
  BS}^z$ in Table~\ref{Tab:SingleCenteredBoostedBH} shows that over
the entire range of boosts considered, $(K R_+/6) P^z$ is within about 
one percent of $P_{\rm BS}^z$.  The ratio of the actual numerical result for
$\Omega_{\rm max}$ to the $K R_+/6$ analytic estimate varies from
$0.931$ to $0.858$ over this range, so it is considerably better to use 
the analytic estimate for $\Omega_{\rm max}$, rather than its actual
numerical value, in relating the Bowen-York boost to the physical 
momentum of a centered black hole.  

The fact that $M_{\rm BS} /M_{\rm irr}$ becomes significantly larger
than $1$ as the black hole motion becomes relativistic shows that the
boosted Bowen-York black holes are definitely {\em not} simply
Schwarzschild black holes moving relative to the asymptotic Minkowski
frame.  There is additional energy associated with non-kinematic
distortions of the geometry.  From Eq.~(\ref{Eq:EnergyBS-A}), we see
that for centered black holes, the direct contribution to the
Bondi-Sachs energy from the conformal extrinsic curvature is
independent of the boost, and for the initial data calculated here,
$(E_{\rm BS})_{\rm K}/M_{\rm irr} \approx - 0.04$, which is small
compared with the positive contribution from conformal factor.  In
contrast, the the Bondi-Sachs momentum is strongly dominated by the
contribution from the conformal extrinsic curvature for centered black
holes, since by Eq.~(\ref{Eq:PvecBS-A}), $(\mathbf P_{\rm BS})_{\rm K}
= (3/K) \bar{\mathbf P}$.

The $Q^i$ boosts are taken to be zero rather than the
inversion-symmetric values often assumed in Bowen-York initial value
calculations on maximal hypersurfaces (e.g., in Ref.~\cite{cook90}).
Only one side of the Einstein-Rosen bridge is astrophysically
relevant, since the other side cannot exert any causal influence on
the astrophysical side.  There is no very compelling reason, other
than geometrical elegance, to assume any symmetry between the two
sides of the bridge or that the bridges of multiple black holes
connect to a common asymptotically flat space.  The geometry of the
physical space in the immediate vicinity of the black holes may depend
to some extent on whether the $Q^i$ are zero or inversion symmetric,
with possible small effects on the results for the Bondi energy and
momentum, but such effects should not be noticeable for 
the near trumpet considered here, and disappear completely in 
the limit of an extreme trumpet configuration $(R_{\rm ms} \to 0)$.

\begin{table*}[ht] 
\begin{tabular}{|c||c|c|c|c|c|c|c|} 
  \hline 
$\bar{P}^z \times 10$ & 0.0 & 0.13889 & 0.27778 & 0.55556 & 
0.83333 & 1.11111 & 1.38889 \\ \hline \hline 
$(K/3)P_{\rm BS}^z \times 10$ & 0.0 & 0.13641 & 0.27352 & 
0.54974 & 0.82708 & 1.10474 & 1.38249 \\ \hline
$(K/3)M_{\rm irr} \times 10$ & 0.28339 & 0.28334 & 0.28338 & 
0.28361 & 0.28382 & 0.28390 & 0.28388 \\ \hline
$P_{\rm BS}^z/M_{\rm irr}$ & 0.0 & 0.48142 & 0.96521 &1.93839 & 
2.91413 & 3.89129 & 4.87005 \\ \hline
$E_{\rm BS}/M_{\rm irr}$ & 1.0 & 1.11221 & 1.41439 & 2.32632 & 
3.39577 & 4.52196 & 5.67844 \\ \hline
$M_{\rm BS}/M_{\rm irr}$ & 1.0 & 1.00261 & 1.03387 & 1.28624 & 
1.74330 & 2.30348 & 2.92016 \\ \hline
 \end{tabular} 
 \caption{\label{Tab:SingleCenteredBoostedBH} 
   Centered black hole with varying Bowen-York boost $\bar{P}^z$.   
   The $\bar{C}$ parameter is fixed at our 
   adopted standard value, and $R_{\rm ms}/R_+$ is adjusted to keep  
   the irreducible mass roughly constant as $\bar{P}^z$ increases.  The 
   input and output quantities are tabulated in the dimensionless 
   and conformal-scale invariant forms discussed earlier in 
   Section~\ref{Sec:4}. }
 \end{table*}

\begin{figure}
 \includegraphics[scale=0.45]{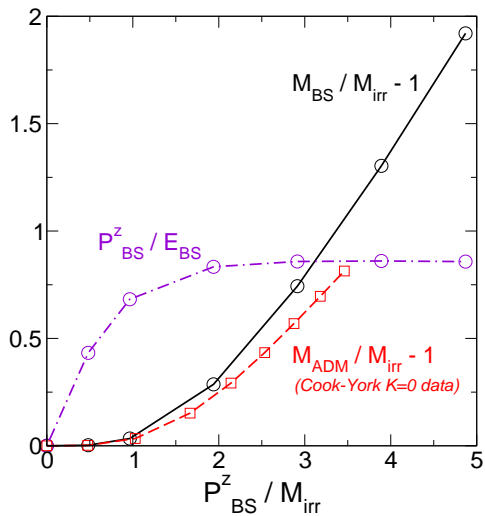}
 \caption{\label{fig:BondiMassForRelativisticP} Centered black hole
   with Bowen-York boosts. Using data from
   Table~\ref{Tab:SingleCenteredBoostedBH} above and Table II of Cook
   and York~\cite{cook90}, we plot (i) the fractional excess of the
   Bondi-Sachs mass over the irreducible mass (solid black line), (ii)
   the corresponding excess ADM mass of Bowen-York initial data on
   maximal hypersurfaces (dashed red curve), and (iii) the velocity of
   the black hole $v^z=P^z_{\rm BS}/E_{\rm BS}$ (dash-dot violet line)
   in our initial data. All quantities are shown as a function of the
   ratio of physical momentum, $P_{\rm BS}$ or $P_{\rm ADM}$ as
   appropriate, to the irreducible mass.  The curves are
   interpolations between results for discrete models.  }
\end{figure}

 For a slightly different slant on our results, we plot in
 Fig.~\ref{fig:BondiMassForRelativisticP} (i) the fractional ``excess
 mass'' of the black hole $M_{\rm BS}/\Mirr - 1$, (ii) the velocity of
 the black hole $v^z \equiv P^z_{\rm BS}/E_{\rm BS}$ in the asymptotic
 Minkowski frame, and (iii) the similar excess ADM mass for a boosted
 Bowen-York black hole on a maximal hypersurface, all as functions of
 the ratio of the appropriate physical momentum (Bondi-Sachs or ADM)
 to the irreducible mass.  Data for the last curve are taken from
 Table II of Ref.~\cite{cook90}, for $C = 0$ and inversion-symmetric
 $Q^i$.  Note that as the boost and physical momentum increase, the
 gravitational mass $M_{\rm BS}$ (solid black curve) increases rapidly
 enough that the black hole's velocity $v^z \equiv P^z_{\rm BS}/E_{\rm
   BS}$ (violet dash-dot curve) asymptotes to a value less than one
 (about $0.86$).  The results of Ref.~\cite{cook90} are quite similar,
 but with a slightly higher asymptotic velocity.  The excess
 gravitational mass can be thought of as energy stored in distortions
 of the geometry relative to a boosted Schwarzschild black hole,
 energy which would be radiated into the black hole and to infinity
 during subsequent evolution.

The black holes on CMC hypersurfaces have more excess mass than the
corresponding black holes on maximal hypersurfaces, modestly so as
long as the motion of the black holes is non-relativistic, but more
substantially as the motion becomes relativistic.  The Cook and York
initial data has $C = 0$, which in the case of spherical symmetry
means that their maximal slice passes through the intersection of the
future and past Schwarzschild event horizons, as opposed to the more
trumpet-like configurations chosen here and emphasized in
Ref.~\cite{Buchman:2009ew}.  How much effect this has on the 
difference in behavior needs further calculations to sort out.

\subsection{Displaced and boosted black holes}
\label{SubSec:SingleDisplacedBoostedBH}
In this section, we lay the groundwork for building a binary black
hole initial data set on CMC hypersurfaces.  For binary black holes in
circular orbit, we want each black hole considered by itself to have
zero radial momentum and transverse momentum corresponding to that of
a mass in a circular orbit, which can be estimated crudely from the
Newtonian equations of motion.  Since we have seen that a displaced
unboosted black hole on a CMC slice has a non-zero radial momentum,
this requires finding the radial boost which makes the net Bondi-Sachs
momentum of a displaced single black hole zero, and then the
transverse boost which generates the desired transverse physical
momentum.  Once this is done for each black hole of the binary
considered separately, the Bowen-York data for the black holes can be
superimposed and the Bondi-Sachs energy and momentum of the full
system calculated.  As long as the two black holes are well-separated,
we should find that the Bondi-Sachs energy of the combined system is
less than the sum of the energies of the black holes considered
separately by something close to the Newtonian binding energy of the
system, provided that the total Bondi-Sachs momentum of the system is
close to zero.

\begin{figure}
  \includegraphics[scale=0.45]{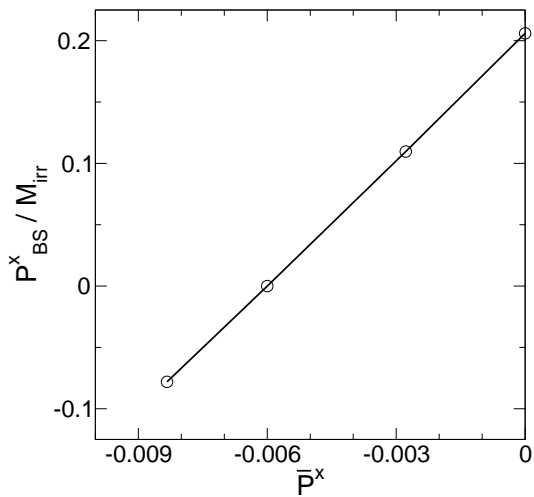}
  \caption{\label{fig:PbyVsPbsParallel} Finding the Bowen-York 
  boost that cancels the Bondi-Sachs momentum associated with 
  the displacement $\bar{D}^x = 0.1$.  The CMC parameters have 
  their usual values, except that $\bar{R}_{\rm ms}$ is varied with 
  the boost to keep $\bar{M}_{\rm irr}$ approximately constant.}
\end{figure}

The first step is to construct a single black hole which is displaced
from the center of the coordinate grid {\em and} at rest.
Fig.~\ref{fig:PbyVsPbsParallel} shows that this result can be achieved
by giving the black hole an appropriate boost antiparallel to its
displacement. For the example shown in
Fig.~\ref{fig:PbyVsPbsParallel}, the hole is displaced in the positive
x-direction, with $\bar{D}^x = 0.1$.  A rescaled boost $\bar{P}^x = -
0.006$ very nearly cancels the radial momentum associated with the
displacement, yielding a net $P^x_{\rm BS}/ \Mirr =
-9.6\times10^{-6}$.

Next, we build an initial data set with two non-spinning displaced
black holes, each of which considered separately has zero Bondi-Sachs
momentum, and calculate the gravitational binding energy; i.e., the
decrease in the Bondi-Sachs energy of the combined system compared
with the sum of Bondi-Sachs energies of each hole separately. The
input parameters are chosen so that, as estimated by the procedures
outlined in Sec.~IIID of~\cite{Buchman:2009ew} for Schwarzschild black
holes, the ratio of minimal surface radius to horizon radius for each
hole is about $0.8$, the ratio of the black hole masses $M^{\rm
  A}/M^{\rm B} \approx 2/1$, and the total rescaled irreducible mass
of the system is about $0.7$ that of our standard single black hole.
We locate holes A and B along the x-axis at $\bar{D}^x_{\rm A} = 0.04$
and $\bar{D}^x_{\rm B} = - 0.08$, so the center of mass of the system
is close to the origin.  The values for $\bar{C}^A$ and $\bar{C}^B$
are, respectively, $0.000228$ and $0.0000568$.  The scaled minimal
surface radii are $\bar{R}_{\rm ms}^{\rm A} = 0.000978$ and
$\bar{R}_{\rm ms}^{\rm B} = 0.000489$.  We apply Bowen-York boosts
which for each black hole in isolation approximately cancel the
Bondi-Sachs radial momentum due to that hole's displacement.  The
dimensionless Bondi-Sachs masses ($\bar{M}_{\rm BS} \equiv
  (K/3)M_{\rm BS}$) for each hole in isolation are $\bar{M}_{\rm
  BS}^A = 0.013104$ and $\bar{M}_{\rm BS}^B = 0.006604$.  When the
black holes are superimposed, we find that the combined Bondi-Sachs
momentum is less than the sum of the already small values for the
individual holes.  The dimensionless Bondi-Sachs mass of the
superimposed black holes is $\bar{M}_{\rm BS} = 0.019368$, which is
$0.000340$ less than the sum of the individual values.  The Newtonian
gravitational binding energy of the system is $ U = M^{\rm A}M^{\rm
  B}/{r_{\rm AB}} \approx \Omega_{\rm max} M^{\rm A}M^{\rm B} /|
\mathbf{D}^A - \mathbf{D}^B|$ or, using the analytic estimate of
$\Omega_{\rm max}$, $K U/3 \approx 0.5 R_+\; \bar{M}_{\rm BS}^{\rm A}
\;\bar{M}_{\rm BS}^{\rm B}/|\bar{\mathbf{D}}^A - \bar{\mathbf{D}}^B | =
0.000371$, which is reasonably close to the rescaled general
relativistic binding energy.

The third step is to determine how Bowen-York boosts transverse to a
black hole's displacement translate into the transverse motion of the
black hole as revealed by its physical transverse momentum.  We have
calculated the Bondi-Sachs energy, linear momentum, and angular
momentum of a displaced black hole with nearly zero radial momentum
and various transverse boosts.  The relationship between transverse
boost and transverse physical momentum is very close to what we found
between boost and momentum for a centered black hole in
Sec.~\ref{SubSec:SingleCenteredBoostedBH}.  An example of the 
comparison for a black hole given a moderately relativistic Bowen-York 
boost transverse to the displacement in the $x$-direction is given in
Table~\ref{Tab:BoostedCntrdVsTransBoostedDispl}.  We conclude that it
is acceptable to use the results for a centered boosted black
hole in Sec.~\ref{SubSec:SingleCenteredBoostedBH}, including the 
analytic formula, when estimating the transverse Bowen-York boosts 
appropriate for black holes in quasi-circular binary orbits.  Of course, 
the displaced black holes also acquire orbital angular momentum 
according to Eq.~(\ref{Eq:AnalAngMom}).  
\begin{table*}[ht] 
\begin{tabular}{|c||c|c|c|c|c|c|} 
  \hline 
   $\bar{D}^x$ & $\bar{P}^y \times 10$ & $(K/3)\Mirr \times 10$ & 
   $P_{\rm BS}^y/\Mirr$ & $E_{\rm BS}/\Mirr$ & $M_{\rm BS}/\Mirr$ & 
   $J^z/\Mirr^2$  \\ \hline \hline
   0 & 0.13889 & 0.28334 & 0.48142&1.11220&1.00261& 0\\ \hline
   0.1 &0.13889 & 0.28336 & 0.47686 &1.11117 &1.00365 & 
   3.45956 \\ \hline
\end{tabular} 
\caption{\label{Tab:BoostedCntrdVsTransBoostedDispl} A centered black
  hole (first row) and a displaced black hole (second row) are 
  both  given the same mildly relativistic (transverse) rescaled 
  Bowen-York boosts $\bar{P}^y =0.013889$.  The displaced black hole 
  is also given a longitudinal rescaled boost $\bar{P}^x= -0.00660$ to 
  approximately cancel the physical momentum 
  associated with the displacement.  The other input parameters for 
  both holes have the standard values.  The black holes have almost 
  identical physical properties, except for the orbital angular momentum 
  of the displaced hole.}  
 \end{table*}
 
 The close relationship between the transverse Bondi-Sachs momentum
 and the transverse rescaled Bowen-York boost can be understood as a
 consequence of the dominance of the direct contribution to the
 Bondi-Sachs momentum from the conformal extrinsic curvature, as
 calculated in Eq.~(\ref{Eq:PvecBS-A}), over the contribution from the
 conformal factor for these transverse components.  As long as
 $\bar{D}$ is reasonably small, the dominant transverse term in
 Eq.~(\ref{Eq:PvecBS-A}) is just $(K R_+/6) P_{\rm trans}$.  There is 
 no simple relation between
 the longitudinal component of the Bondi-Sachs momentum and the
 longitudinal component of the Bowen-York boost because the conformal
 factor contribution to the longitudinal component is not small.

\subsection{Binary black hole initial data}
\label{SubSec:BBhId}

Here we build on the results of the previous subsections to construct
initial data for two spinning unequal-mass black holes with initial
momenta roughly appropriate for circular orbits.  The input parameters
are chosen based on a Newtonian analysis, in circumstances such that
the post-Newtonian corrections are roughly at the $10 \%$ level.
Ignoring relativistic corrections is consistent with other sources of
uncertainty, such as estimating parameters for each black hole in
isolation, ignoring their tidal interactions.  While similar initial
data were presented in~\cite{Buchman:2009ew}, we are now on much
firmer ground in considering the relationship
between the Bowen-York parameters and the physical properties of each
hole.

We consider two examples.  Both are constructed to have about a 2:1
mass ratio and roughly a 12:1 ratio of physical distance between the 
black holes to the total gravitational mass of the system.  The minimal
surface radii are chosen to give moderately trumpet-like
configurations.  The black holes are assigned close to maximal spins
oriented in the orbital plane and perpendicular to the displacement
between the holes.  In one case (Case I), the ratio of coordinate
separation of the holes to the coordinate radius of $\ScriPlus$ is
$0.12$, and in the other case (Case II), it is three times larger.  In
the latter case, there is the potential for greater computational
efficiency, since the black holes can be resolved with a lower
resolution grid.

The Newtonian analysis of a circular orbit binary system with masses
$M_{\rm A}$ and $M_{\rm B}$ separated by a distance $r_{\rm AB}$ gives
for the equal and opposite momenta of the masses
\[
P_{\rm A}^y = - P_{\rm B}^y = \frac{M_{\rm A} M_{\rm B}}
{M_{\rm A} + M_{\rm B}} \sqrt{\frac{M_{\rm A} + M_{\rm B}}{r_{\rm AB}}},
\]
perpendicular to the displacements of the black holes (which are along
the $x$-axis).  Equating the transverse rescaled Bowen-York boost to
$K/3$ times the transverse physical momentum, for $M_{\rm A} \approx 2
M_{\rm B}$, we choose $\bar{P}^y = 0.064(K/3) (M_{\rm A} + M_{\rm
  B})$, since $r_{\rm AB}$ is about $12$ times the total mass.
Estimating black hole masses from the Bowen-York parameters is a bit
uncertain when the black holes are not Schwarzschild, but consistency
with the total system mass can be verified retrospectively.

The non-zero input parameters chosen for Case I
are $\bar{C}_{\rm A} = 2.4531\times 10^{-5}$, $\bar{C}_{\rm B} =
5.1224 \times 10^{-6}$, $\bar{S}_{\rm A}^y = 6.5224$, $\bar{S}_{\rm
  B}^y = -7.8088$, $\bar{D}_{\rm A}^x = 0.04$, $\bar{D}_{\rm B}^x =
-0.08$, $\bar{R}_{\rm ms}^{\rm A} = 9.78 \times 10^{-4}$,
$\bar{R}_{\rm ms}^{\rm B} = 4.89 \times 10^{-4}$, and boost components
$\bar{P}_{\rm A}^x = -0.001074$, $\bar{P}_{\rm A}^y = 0.00116$,
$\bar{P}_{\rm B}^x = 0.001060$, $\bar{P}_{\rm B}^y = -0.00116$.  As
noted in Sec.~\ref{SubSec:SingleSpinningBh}, we expect the rescaled
black hole masses to scale roughly as $\bar{C}_{\rm eff} = \left[
  \bar{C} \sqrt{1 + 2 \left(\bar{S}\right)^2} \right]^{1/2}$, which is
$0.0150864$ for hole A and $0.0075366$ for hole B.  The large values
of $\bar{S}$ for each hole mean that the black holes have close to the
maximum possible spin extremality parameter $\zeta$.

In Case II, the input parameters are $\bar{C}_{\rm A} = 3.6231 \times
10^{-4}$, $\bar{C}_{\rm B} = 8.5343 \times 10^{-5}$, $\bar{S}_{\rm
  A}^y = 3.9347$, $\bar{S}_{\rm B}^y = -4.1761$, $\bar{D}_{\rm A}^x =
0.12$, $\bar{D}_{\rm B}^x = -0.24$, $\bar{R}_{\rm ms}^{\rm A} = 2.934
\times 10^{-3}$, $\bar{R}_{\rm ms}^{\rm B} = 1.467 \times 10^{-3}$,
and boost components $\bar{P}_{\rm A}^x = -0.010026$, $\bar{P}_{\rm
  A}^y = 0.00348$, $\bar{P}_{\rm B}^x = 0.010617$, $\bar{P}_{\rm B}^y
= -0.003639$.  The $\bar{C}_{\rm eff}$ values are $0.04526$ for hole A
and $0.02261$ for hole B, about three times the values in Case I,
while the $\bar{D}$ values are also three times larger.  This
suggests the two cases are similar in the ratio of physical separation
of the holes to the total mass of the system.  The $\bar{S}$
 values are smaller in Case II than in Case I; this is because the
elliptic solver failed to converge if we tried to make
the $\bar{S}$ values as large as in Case I.  Another effect of the
larger ratio of the size of the binary system to $3/K$ in Case II is
the larger $\bar{P}^x$ values needed to cancel out the radial
components of the individual black hole momenta.

Our results for the irreducible masses and angular momenta of the
individual holes and the physical energy, momentum, mass,
and angular momentum of the systems are given in Table~\ref{Tab:BBH}.
Note that there are non-zero, though small, $z$-components for the
system velocities even though none of the boosts has
a non-zero $z$-component.  This is the result of the spins of the
black holes interacting with the geometry of the hypersurface, one
indication of which is the $\mathbf{D} \times \mathbf{S}$ term in
Eq.~(\ref{Eq:PvecBS-A}).  Also, note that $K/3 = 0.02$ gives a total
system mass close to $1$ in Case I, and this together with $R_+ = 100$
makes coordinate distances close to physical distances.  The same is
true for Case II since $K$ is increased and $R_+$ is decreased by a
factor of $3$.  The calculations for Case II took about $80\%$ of the
time on the same computer as the calculations for Case I, indicating
some potential for realizing increased efficiency of computation by
having the binary system occupy a larger fraction of the conformal
domain.

\begin{table*}[ht] 
\begin{tabular}{|r||c|c|c|c|c|c|c|c|c|} 
  \hline 
  ~~&$\bar{M}_{\rm irr}$
  &$(K/3) E_{\rm BS}$
  &$(K/3) M_{\rm BS}$
  &$P_{\rm BS}^x/E_{\rm BS}$
  &$P_{\rm BS}^y/E_{\rm BS}$
  &$P_{\rm BS}^z/E_{\rm BS}$
  &$(K/3)^2 J^x$
  &$(K/3)^2 J^y$
  &$(K/3)^2 J^z$ \\
  ~~&~$\times 10^2$
  &~~~~~$\times 10^2$
  &~~~~~$\times 10^2$
  &~~~~~$\times 10^3$
  &~~~~~$\times 10^3$
  &~~~~~$\times 10^3$
  &~~~~$\times 10^4$
  &~~~~$\times 10^4$
  &~~~~$\times 10^4$ 
\\ \hline \hline
  \bf{Case I}&&&&&&&&&\\ \hline
  {\em hole A:~}&1.06367&n/a&n/a&n/a&n/a&n/a&0&~1.60000&~0.92800 \\ \hline
  {\em hole B:~}&0.53449&n/a&n/a&n/a&n/a&n/a&0&-0.40000 &~1.85600 \\ \hline
  {\em system:~}&1.59816&1.96685&1.96685&-1.04379&~0.25464&0.05850&0
  &~1.20000&~2.78400 \\ \hline
  \bf{Case II}&&&&&&&&& \\ \hline
  {\em hole A:~}&3.16247&n/a&n/a&n/a&n/a&n/a&0&14.25600&~8.35200 \\ \hline
  {\em hole B:~}&1.62032&n/a&n/a&n/a&n/a&n/a&0&-3.56400&17.46720 \\ \hline
  {\em system:~}&4.78280&5.90011&5.90003&-5.25838&-0.16546&0.17739&0
  &10.69200&25.81920 \\ \hline \hline
\end{tabular} 
\caption{\label{Tab:BBH}  
  Physical quantities for two binary black hole
  initial data sets. (See text for input parameter values.) From left
  to right are the irreducible masses, the Bondi-Sachs energy, the 
  Bondi-Sachs mass, the system velocity components, and the angular 
  momentum components, all made dimensionless by appropriate 
  powers of $K/3$.  There are no well-defined Bondi-Sachs energy and 
  linear momentum for the individual black holes, but the angular momenta 
  can be calculated using Eq.~(\ref{AppEq:AnalAngMom}).
}
\end{table*}

While the system momentum and velocity should ideally be zero, this is
not precisely the case in our calculations, which made no attempt to
fine-tune the input parameters.  Still, the residual system velocities
are quite small compared with the Newtonian estimates for the orbital
velocities.  The system mass is significantly greater than the sum of
the irreducible masses in both Cases I and Case II, since the
contributions of the spins to the black hole hole masses dominates
over their gravitational binding energy.  The slightly smaller spins
in Case II make this effect somewhat smaller there.  Of course, the
system mass can never be less than the total irreducible mass.

\section{Discussion}

\label{Sec:Discussion}

The first part of this paper is a discussion of the asymptotic
behavior of the conformal factor $\Omega$, the conformal spatial
metric, and the conformal extrinsic curvature of CMC hypersurfaces in
the neighborhood of $\ScriPlus$ as constrained by the initial value
equations for astrophysically general asymptotically flat spacetimes,
and a calculation of the Bondi-Sachs mass aspect in terms of the
coefficients in the expansion away from the $\dot{\ScriPlus}$
2-surface, which is the intersection of the CMC hypersurface with the
$\ScriPlus$ null hypersurface.  In this analysis, we adopt special
spatial coordinates, Gaussian normal coordinates based on
$\dot{\ScriPlus}$, and use the conformal proper distance $z$ from
$\dot{\ScriPlus}$ along the normal spatial geodesics as the expansion
parameter.  Angular coordinates are propagated along the normal
spatial geodesics, initialized to be standard polar coordinates on
$\dot{\ScriPlus}$, which is assumed to be a true sphere with intrinsic
curvature $\xi_0^2$.  This is not a physical constraint for an
asymptotically flat spacetime with a topologically spherical null
infinity, just a constraint on the conformal gauge.  The asymptotic
expansion is essentially identical to that presented in
Ref.~\cite{Andersson1994}, except we do not allow the polyhomogeneous
terms in the expansion of the angular part of the conformal spatial
metric through $O(z^2)$ that are a major focus of that paper.  No such 
terms are present unless they are present at some arbitrarily early 
time, which means that they should be associated with 
incoming radiation at past null infinity.  In typical astrophysical 
problems incoming radiation near $\ScriPlus$ should only be 
generated by backscatter or nonlinear interaction of outgoing 
radiation, and should not be present {\em ab initio}, except for what 
might be plausibly associated with backscatter of radiation emitted 
before the start of the numerical calculation.

As noted in Ref.~\cite{Andersson1994}, polyhomogeneous terms are
nevertheless generically present at $O(z^4)$ in the expansion of the
conformal factor and at $O(z^2)$ in the expansion of the conformal
extrinsic curvature.  The authors of Ref.~\cite{Andersson1994}
considered the vanishing of the expression in Eq.~(\ref{Eq:Qdefb}) a
condition for the absence of these polyhomogeneous terms, but we point
out that the condition cannot be satisfied over any finite time
interval in which outgoing gravitational radiation, as indicated by
time variation of $\breve{\chi}_{\rm AB}$, is present at $\ScriPlus$
(see also Ref.~\cite{Bardeen2011a}).  The implied smoothness of only
$C^3$ for the conformal factor and $C^1$ for the conformal extrinsic
curvature at $\ScriPlus$ is a gauge artifact of the CMC hypersurface
condition.  The offending terms are not present if the physical mean
curvature $K$ has precisely the right angular dependence at $O(z^3)$
in a power series expansion away from $\dot{\ScriPlus}$.  This cannot
be accomplished by any predetermined specification of $K$, but
maintaining a high degree of smoothness at $\ScriPlus$ is possible
with fully hyperbolic evolution schemes, such as the regular
conformal field equations of Friedrich~\cite{Friedrich1983,
  Friedrich1986}, as well as with the Bondi-Sachs gauge.

For the assumed special geometric properties of $\dot{\ScriPlus}$ to
be preserved during evolution, a constraint must be placed on the
evolution of the conformal factor; specifically, that the trace of the
conformal extrinsic curvature $\tilde{K}$ satisfy the boundary
condition of Eq.~(\ref{Eq:Ktildebc}).  This says that the 3D mean
conformal extrinsic curvature of the CMC hypersurface at
$\dot{\ScriPlus}$, which governs the evolution of the conformal
factor, should equal the 2D mean extrinsic curvature of the
$\dot{\ScriPlus}$ 2-surface when both are renormalized according to
their respective number of dimensions.  All this also assumes
preservation of the coordinate location of $\ScriPlus$ as a coordinate
sphere centered at the origin and propagation of the angular
coordinates on $\ScriPlus$ along its null generators, which imposes
boundary conditions on the shift vector.  These boundary conditions
can be implemented in a mixed hyperbolic-elliptic evolution scheme
like those of Ref.~\cite{Moncrief2009} or Ref.~\cite{Bardeen2011a}.
Also, we impose the usual regularity conditions at $\ScriPlus$, the
``zero-shear'' condition and the vanishing of the Weyl tensor, which
as shown for the evolution schemes of
\cite{Moncrief2009,Bardeen2011a}, are preserved by the evolution
equations.

We have applied this asymptotic analysis to derive the expression for
the Bondi-Sachs mass aspect given in Eq.~(\ref{Eq:MassAspect}), from
which the Bondi-Sachs energy and momentum can be calculated as
appropriate angular averages over $\dot{\ScriPlus}$.  This is a 3D
expression, involving only quantities evaluated on a single CMC
hypersurface.  The 3D expression derived in Section 5.3 of
Ref.~\cite{Chrusciel2004} requires knowledge of a ``background''
metric and extrinsic curvature in addition to the actual metric and
extrinsic curvature of the CMC hypersurface.  It is non-trivial to
determine the appropriate background quantities.  The terms appearing
in Eq.~(\ref{Eq:MassAspect}) are simply related to the asymptotic
behavior of the conformal factor, conformal metric, and conformal
extrinsic curvature in Gaussian normal coordinates, and can be
determined by solving the geodesic equation for the geodesics normal
to $\dot{\ScriPlus}$ in the CMC hypersurface in any computational
spatial gauge satisfying our boundary conditions.  The alternative
expression of Eq.~(\ref{Eq:Weylformula}) based on the Weyl tensor is
completely equivalent, as can be seen by evaluating the Weyl tensor in
the form of Eq.~(\ref{Eq:WeylE}) in Gaussian normal coordinates.

With conformally flat initial data, all of this simplifies drastically.  With
$\dot{\ScriPlus}$ a coordinate sphere of constant radius in a flat
conformal geometry, the normal geodesics are radial, and the 2D
extrinsic curvature of surfaces of constant coordinate radius $R$ is
just $R^{-1}$ times the unit sphere metric.  The momentum constraint
equation admits solutions identical in form to the Bowen-York
solutions.  We have shown in the Appendices how to evaluate
analytically the angular integrals of the term coming from the
conformal extrinsic curvature in the mass aspect in terms of the
Bowen-York parameters.  The only numerical calculation necessary is
solving the Hamiltonian constraint equation for the conformal factor
and fitting to a power series expansion in the coordinate distance
from $\dot{\ScriPlus}$, taking advantage of the known analytic form
through order $z^5$.  We have computed how the Bondi-Sachs
energy-momentum and the angular momentum of the system depend on the
Bowen-York parameters for a sampling of single and double black hole
systems, similar to those considered in Ref.~\cite{Buchman:2009ew}.

For a single black hole centered at the origin, with varying
Bowen-York spin (which in this case equals the system angular
momentum), the results are displayed in Fig.~\ref{fig:Diff_BE_Mc}.
They show that the ``rotational energy'' of the black hole, the excess
of the Bondi-Sachs mass over the irreducible mass, has a dependence on
the spin extremality parameter $\zeta$ similar to that of Kerr black
holes.  The Bondi-Sachs mass is a bit larger in our case, as one would
expect.  The main difference is that our sequence cannot be extended
to as large a value of the extremality parameter as the Kerr sequence.
 
A single non-spinning, unboosted black hole displaced from the origin,
i.e., displaced relative to the center of the sphere representing
future null infinity, is more interesting.  As shown in
Fig.~\ref{fig:CxVsPx}, the displaced black hole acquires a velocity in
the asymptotic Minkowski frame in the direction of its displacement
close to twice the ratio of the displacement to $R_+$. Consequently,
when constructing initial data for black hole binaries on CMC
hypersurfaces, compensating Bowen-York boosts directed inward should
be introduced for each displaced black hole, so that its net radial
velocity is zero.
 
Results for a single centered black hole given various, mostly rather
large, boosts are presented in Sec.~\ref{SubSec:SingleDisplacedBh}.
Here we emphasize how close the numerical result for the Bondi-Sachs
momentum $\mathbf{P}_{\rm BS}$ is to $KR_+/6$ times the Bowen-York
boost vector, even when the black hole is moving at relativistic
speeds.  Ref.~\cite{Buchman:2009ew}, presented a crude hand-waving
argument that the physical momentum should be roughly $\Omega_{\rm
  max}$ (calculated numerically) times the boost.  The expected
accuracy of this estimate was of order $K M_{\rm irr}$, or about $10
\%$ for our choices of Bowen-York parameters.  Comparison of the first
two rows of Table~\ref{Tab:SingleCenteredBoostedBH} shows that $K
R_+/6$ times the boost is within about $1\%$ of $\mathbf{P}_{\rm BS}$,
while $K R_+/6$ differs from $\Omega_{\rm max}$ by about $10 \%$.  The
physical properties of these boosted centered black holes are explored
further in Fig.~\ref{fig:BondiMassForRelativisticP} and compared with
corresponding results for boosted black holes on maximal
hypersurfaces.  In both cases, the ratio of physical mass to
irreducible mass increases rather rapidly with increasing boost, and
the velocity of the black hole plateaus at a value somewhat less than
the speed of light for large boosts.  Bowen-York initial data cannot
produce black holes moving at speeds arbitrarily close to the speed of
light.
 
A combination of boosts and displacements is necessary to construct
initial data for binary black holes.  In
Section~\ref{SubSec:SingleDisplacedBoostedBH}, we first determine the
radial boost opposite to the direction of the displacement necessary
to cancel the radial momentum induced by the displacement.  Orbital
motion of a black hole in a binary system requires a transverse boost,
and we verified in Table~\ref{Tab:BoostedCntrdVsTransBoostedDispl}
that the relationship between the {\em transverse} boost and {\em
  transverse} momentum for a displaced black hole is very nearly the
same as the relation between boost and momentum for a centered boosted
black hole, as calculated in Section~\ref{SubSec:SingleDisplacedBh}.
We also checked that the gravitational binding energy of two black
holes at rest separated by about 10 times the total gravitational mass
of the system agrees reasonably well with the Newtonian formula.
 
Finally, we make a crude attempt to construct initial data for two
black holes in circular orbits about the center of mass.  Again, the
black holes, with about a 2:1 mass ratio, are separated by a proper
distance about 10 times the gravitational mass of the system.  Boosts
are assigned so that, considered individually, each black hole has
zero radial momentum and transverse momentum roughly as expected from
a Newtonian analysis of the orbit.  The black holes are also given
close to maximal spins oriented in the plane of the orbit.  The total
energy, momentum, and angular momentum of the system are calculated
and displayed in Table~\ref{Tab:BBH}.  The numerical results are
consistent with the system linear momentum being close to zero, in
that the system $P^i_{\rm BS}/E_{\rm BS}$ values are rather small
compared to the $\bar{P}^i/\bar{M}_{\rm irr}$ for each black hole.
The component of the residual momentum along the line of the black
hole displacements (the $x$-component) is considerably larger than the
other components, suggesting that we should have taken account of the
black hole spins in estimating the radial boosts.  The system angular
momentum, including both spin and orbital angular momentum, is roughly
consistent with Newtonian expectations.  The two cases we present are
designed to have similar ratios of physical separation to black hole
masses, but values of $(K/3)M_{\rm BS}$ differing by about a factor of
three.  The ratio of the coordinate separation of the holes to $R_+$
is three times larger in Case II than in Case I, implying
correspondingly larger effects of the warping of the CMC hypersurface
at the scale of the binary system.  Nonetheless, Case II achieves
close to the same accuracy as Case I with half the number of
collocation points in the outer spherical computational domain
surrounding the two holes, $40$ versus $80$.
 
The exploratory analyses of conformally flat initial data on CMC
hypersurfaces carried out in Ref.~\cite{Buchman:2009ew} and this paper
need to be extended in various ways to approach the current
sophistication of initial data calculations on maximal hypersurfaces.
In particular, it would be interesting to see how well the numerical
techniques used here hold up when considering non-conformally-flat
data, particularly when radiation is present at $\dot{\ScriPlus}$.
How well does the {\tt SpEC} elliptic solver handle the more
complicated asymptotic behavior, including the polyhomogeneous terms
in the conformal factor and extrinsic curvature?  This is important
for implementing mixed hyperbolic-elliptic evolution systems.  Also,
quasi-equilibrium methods~\cite{Cook2002, Cook2004,Grandclement2001}
based on the conformal-thin-sandwich approach to the initial data
problem~\cite{York1999, Pfeiffer2003b} have come to the fore in recent
years as a way of obtaining quieter starts to binary black hole
calculations, with less junk radiation and more circular initial
orbits.  It is clearly worthwhile to attempt to adapt these approaches
to initial data on CMC hypersurfaces.  Can initial data for more
Kerr-like black holes, as in Ref.~\cite{Lovelace2008}, be constructed
on CMC hypersurfaces?  We hope to explore such issues in future
papers.


\begin{acknowledgments}
  We gratefully acknowledge Harald Pfeiffer for helpful feedback, and
  for advice regarding the {\tt SpEC} elliptic solver. JMB thanks the
  Perimeter Institute for their hospitality during the final stages of
  writing the paper.
\end{acknowledgments}

\begin{appendix}
  \section{Analytic calculation of the partial Bondi-Sachs
    four-momentum from the conformal extrinsic curvature}
\label{AppA}

The Bondi-Sachs mass aspect for the conformally flat initial data is given in
Eq.~(\ref{Eq:BSMassAspect}) as the sum of two terms, one a coefficient
in the expansion of the conformal factor away from $\dot{\ScriPlus}$, 
and the other a projection $\bar{A}_{RR}$ of the conformal extrinsic
curvature defined by Eq.~(\ref{Eq:ARRdef}).  The first term requires
solving a non-linear elliptic equation, which can only be done
numerically.  However, in the case of Bowen-York initial data on a
conformally flat spatial geometry, the conformal extrinsic curvature
$\tilde A_{ij} $ is as given explicitly in
Eq.~(\ref{Eq:GeneralizedBY}).  The angular integrals on $\dot{\ScriPlus}$
for the $\bar{A}_{RR}$ contribution to the Bondi-Sachs mass and linear
momentum will be obtained in analytic form with the help of {\tt
  Mathematica}.

For a given black hole, we can rotate the Cartesian coordinates in the
conformal flat space to locate the black hole on the polar (positive
$z$-) axis.  The unit vector $\mathbf N$ normal to $\dot{\ScriPlus}$
has Cartesian components $N^i = \left( {\sin \theta \cos \phi , \,
    \sin \theta \sin \phi, \, \cos \theta } \right)$.  The
displacement from the black hole location to a point on
$\dot{\ScriPlus}$ has components
\[
r^i = \left( {R_ + \sin \theta \cos \phi , \, R_ + \sin \theta \sin
    \phi, \, R_ + \cos \theta - D} \right) \equiv rn^i ,
\]
where $\mathbf n$ is a unit vector.  With $q \equiv D / R_+$ and $x
\equiv \cos \theta $, the coordinate distance from the black hole to
the point on $\dot{\ScriPlus}$ is
\[
r = R_ +  \sqrt {1 + q^2  - 2qx} .
\]
The scalar and vector products of the two unit vectors are
\[
\mathbf N \cdot \mathbf n = \frac{{1 - qx}} {{\sqrt {1 + q^2 - 2qx}
  }},\,\,\,\,\,\,\mathbf N \times \mathbf n = \frac{{q\sin \theta }} {{\sqrt
    {1 + q^2 - 2qx} }}\left( { - \sin \phi, \, \cos \phi, \, 0}
\right).
\]

The direct contribution to the Bondi-Sachs mass aspect $M_{\rm A}$
from the conformal extrinsic curvature is
\begin{equation}
\label{Eq:MassBS-A}
\left( {M_{{{\rm A}}} } \right)_{{\rm K}} = - \frac{3} {K}
\bar{A}_{RR} = - \frac{K} {6}R_ + ^3 N^k \tilde A_{k \ell} N^{\ell} .
\end{equation}
Substituting the expression in Eq.~(\ref{Eq:GeneralizedBY}) gives 
\begin{eqnarray}
  \left( {M_{{{\rm A}}} } \right)_{{\rm K}} &=& - \frac{{KC}}
  {6}\left( {\frac{R} {r}} \right)^3 \left[ {3\left( {\mathbf N \cdot
          \mathbf n} \right)^2 - 1} \right] - K\left( {\frac{R}
      {r}} \right)^3 \mathbf S \cdot \left( {\mathbf N \times \mathbf n}
  \right)\left( {\mathbf N \cdot \mathbf n} \right) \nonumber \\
  &&+ \frac{{KR_ + }} {4}\left( {\frac{{R_ + }}
      {r}} \right)^2 \left[ {2\left( {\mathbf N \cdot \mathbf P}
      \right)\left( {\mathbf N \cdot \mathbf n} \right) + \left( {\mathbf n
          \cdot \mathbf P} \right)\left( {\left( {\mathbf N \cdot \mathbf n}
          \right)^2  - 1} \right)} \right] \nonumber \\
  &&- \frac{K} {{4R_ + }}\left( {\frac{{R_ + }}
      {r}} \right)^4 \left[ {2\left( {\mathbf N \cdot \mathbf Q} \right)
      \left( {\mathbf N \cdot \mathbf n} \right) + \left( {\mathbf n \cdot \mathbf Q}
      \right)
      \left( {1 - 5\left( {\mathbf N \cdot \mathbf n} \right)^2 } \right)}
  \right]. \nonumber
\end{eqnarray} 

The contribution of this to the Bondi-Sachs energy is the average of
Eq.~(\ref{Eq:MassBS-A}) over solid angle.  Since the average of any
term with an odd power of $\cos \phi$ or $\sin \phi$ vanishes, the
spin term and the components of $\mathbf P$ and $\mathbf Q$
perpendicular to $\mathbf D$ do not contribute.  We have
\begin{eqnarray}
  \left( {E_{{{\rm BS}}} } \right)_{{\rm K}} &=& - \frac{{KC}}
  {6}\left[ {\frac{3} {2}\int\limits_{ - 1}^1 {dx\frac{{\left( {1 -
                qx} \right)^2 }} {{\left( {1 + q^2 - 2qx}
            \right)^{5/2} }} - \frac{1} {2}\int\limits_{ - 1}^1
        {\frac{{dx}}
          {{\left( {1 + q^2  - 2qx} \right)^{3/2} }}} } } \right]
  \nonumber \\
  &&+ \frac{{KR_ + }} {4}P^z \left[ {\int\limits_{ - 1}^1
      {dx\frac{{x\left( {1 - qx} \right)}} {{\left( {1 + q^2 - 2qx}
            \right)^{3/2} }} + \frac{1} {2}\int\limits_{ - 1}^1
        {dx\frac{{\left( {x - q} \right)\left( {1 - qx} \right)^2 }}
          {{\left( {1 + q^2 - 2qx} \right)^{5/2} }} - \frac{1}
          {2}\int\limits_{ - 1}^1 {dx\frac{{\left( {x - q} \right)}}
            {{\left( {1 + q^2  - 2qx} \right)^{3/2} }}} } } } \right]
  \nonumber \\
  &&- \frac{K} {{4R_ + }}Q^z \left[ {\int\limits_{ - 1}^1
      {dx\frac{{x\left( {1 - qx} \right)}} {{\left( {1 + q^2 - 2qx}
            \right)^{5/2} }} + \frac{1} {2}\int\limits_{ - 1}^1
        {dx\frac{{\left( {x - q} \right)}} {{\left( {1 + q^2 - 2qx}
              \right)^{5/2} }} - \frac{5} {2}\int\limits_{ - 1}^1
          {dx\frac{{\left( {x - q} \right)\left( {1 - qx} \right)^2 }}
            {{\left( {1 + q^2  - 2qx} \right)^{7/2} }}} } } } \right].\nonumber
\end{eqnarray}
Evaluating the integrals with {\tt Mathematica},
\begin{eqnarray}
  \left( {E_{{{\rm BS}}} } \right)_{{\rm K}} &=& - \frac{{KC}}
  {6}\left[ {\frac{{3 - 2q^2 }} {{1 - q^2 }} - \frac{1} {{1 - q^2 }}}
  \right] + \frac{{KR_ + }} {4}P^z \left[ {\frac{4}
      {3}q + 0 - 0} \right] \nonumber \\
&&- \frac{K} {{4R_ + }}Q^z \left[ {\frac{4} {3}\frac{{2q
          - q^3 }} {{\left( {1 - q^2 } \right)^2 }} + \frac{2}
      {3}\frac{q} {{\left( {1 - q^2 } \right)^2 }} - \frac{2}
      {3}\frac{{5q - 2q^3 }}
      {{\left( {1 - q^2 } \right)^2 }}} \right], \nonumber
\end{eqnarray} 
which simplifies to
\begin{equation}
\label{Eq:EnergyBS-A}
\left( {E_{{{\rm BS}}} } \right)_{{\rm K}} = - \frac{{KC}} {3} +
\frac{K} {3}\left( {\mathbf D \cdot \mathbf P} \right) = 
\frac{3}{K} \left[ - \bar{C} + 2  \bar{\mathbf D} \cdot 
\bar{\mathbf P} \right].
\end{equation}

The Bondi-Sachs momentum $\mathbf P_{{{\rm BS}}} $ is the average over solid
angle of $M_{{{\rm A}}} \,\mathbf N$ .  The portion of this directly due
to the extrinsic curvature is best calculated separately for the
components parallel and perpendicular to the displacement $\mathbf D$ of
the black hole.  The component parallel to $\mathbf D$ is obtained simply
by inserting a factor of $x = \cos \theta$ into each of the integrals
in the above expression for $ \left( {E_{{{\rm BS}}} }
\right)_{{\rm K}}$ .  Evaluating the integrals gives
\begin{eqnarray}
  \left( {P_{{{\rm BS}}}^z } \right)_{{\rm K}} &=& - \frac{{KC}}
  {6}\left[ {\frac{{3q - 2q^3 }} {{1 - q^2 }} - \frac{q} {{1 - q^2 }}}
  \right] + \frac{{KR_ + }} {4}P^z \left[ {\frac{{10 + 12q^2 }} {{15}}
      + \frac{{5 - 2q^2 }} {{15}} - \frac{1}
      {3}} \right] \nonumber \\
  && - \frac{K} {{4R_ + }}Q^z \left[ {\frac{2} {3}\frac{{1 +
          3q^2 - 2q^4 }} {{\left( {1 - q^2 } \right)^2 }} + \frac{1}
      {3}\frac{{1 + q^2 }} {{\left( {1 - q^2 } \right)^2 }} - \frac{1}
      {3}\frac{{5 + 3q^2 - 2q^4 }}
      {{\left( {1 - q^2 } \right)^2 }}} \right], \nonumber
\end{eqnarray} 
which simplifies to
\begin{equation}
\label{Eq:PzBS-A}
    \left( {P_{{{\rm BS}}}^z } \right)_{{\rm K}} = - \frac{{KC}}
    {3}\frac{D} {{R_ + }} + \frac{{KR_ + }} {6}P^z \left( {1 + \left(
    {\frac{D} {{R_ + }}} \right)^2 } \right) + \frac{K}{6 R_ + }  
    Q^z .
\end{equation}
The perpendicular component has a contribution from the spin term in
the extrinsic curvature.  For instance, the $x$-component is
\begin{eqnarray}
  \left( {P_{{{\rm BS}}}^x } \right)_{{\rm K}} &=& \frac{{KR_ + }}
  {8}P^x \left[ {\int\limits_{ - 1}^1 {dx\frac{{\left( {1 - x^2 }
            \right)\left( {1 - qx} \right)}} {{\left( {1 + q^2 - 2qx}
            \right)^{3/2} }} + \frac{1} {2}\int\limits_{ - 1}^1
        {dx\frac{{\left( {1 - x^2 } \right)\left( {1 - qx} \right)^2
            }} {{\left( {1 + q^2 - 2qx} \right)^{5/2} }} - \frac{1}
          {2}\int\limits_{ - 1}^1 {dx\frac{{\left( {1 - x^2 }
                \right)}}
            {{\left( {1 + q^2  - 2qx} \right)^{3/2} }}} } } } \right]
  \nonumber \\
  && - \frac{K} {{8R_ + }}Q^x \left[ {\int\limits_{ - 1}^1
      {dx\frac{{\left( {1 - x^2 } \right)\left( {1 - qx} \right)}}
        {{\left( {1 + q^2 - 2qx} \right)^{5/2} }} + \frac{1}
        {2}\int\limits_{ - 1}^1 {dx\frac{{\left( {1 - x^2 } \right)}}
          {{\left( {1 + q^2 - 2qx} \right)^{5/2} }} - \frac{5}
          {2}\int\limits_{ - 1}^1 {dx\frac{{\left( {1 - x^2 }
                \right)\left( {1 - qx} \right)^2 }}
            {{\left( {1 + q^2  - 2qx} \right)^{7/2} }}} } } } \right]
  \nonumber \\
  && - \frac{{KD^z S^y }} {{4R_ + }}\int\limits_{ - 1}^1
  {dx\frac{{\left( {1 - x^2 } \right)\left( {1 - qx} \right)}}
    {{\left( {1 + q^2  - 2qx} \right)^{5/2} }}} . \nonumber
\end{eqnarray}
Evaluating the integrals gives 
\begin{eqnarray}
  \left( {P_{{{\rm BS}}}^x } \right)_{{\rm K}} &=& \frac{{KR_ + }}
  {8}P^x \left[ {\frac{{20 - 12q^2 }} {{15}} + \frac{{10 - 8q^2 }}
      {{15}} - \frac{2} {3}} \right] - \frac{K} {{8R_ + }}Q^x \left[
    {\frac{4} {3} + \frac{2} {3}\frac{1} {{1 - q^2 }} - \frac{2}
      {3}\frac{{5 - 4q^2 }} {{1 - q^2 }}} \right] - \frac{{KD^z S^y }}
  {{3R_ +  }} \nonumber \\
  &=& \frac{{KR_ + }} {6}P^x \left[ {1 - \left( {\frac{D} {{R_ +
            }}} \right)^2 } \right] + \frac{K} {{6R_ + }}Q^x -
  \frac{{KD^z S^y }}
  {{3R_ +  }}. \nonumber
\end{eqnarray}
Combine this result with Eq.~(\ref{Eq:PzBS-A}) to get in vector form
\begin{equation}
\label{Eq:PvecBS-A}
    \left( {\mathbf P_{{{\rm BS}}} } \right)_{{\rm K}} = \frac{3}{K} \left[ 
    - \bar{C} \bar{\mathbf D} + \bar{\mathbf P} \left( 1 - \left| \bar{\mathbf D} 
    \right|^2 \right) + 2 (\bar{\mathbf D} \cdot \bar{\mathbf P}) \bar{\mathbf D} 
    + \bar{\mathbf Q} + \bar{C} \bar{\mathbf D} \times \bar{\mathbf S} 
    \right].
 \end{equation}

The expressions in Eqs.~(\ref{Eq:EnergyBS-A}) and~(\ref{Eq:PvecBS-A})
are only portions of the total Bondi-Sachs energy and momentum, 
respectively, so nothing can be concluded from these
expressions about how the total Bondi-Sachs energy and momentum 
depend on the Bowen-York parameters.  In
particular, the total Bondi-Sachs energy is always positive even when
the typically dominant first term in Eq.~(\ref{Eq:EnergyBS-A}) is
negative.  Also, note that a displaced ``Schwarzschild'' black hole
(with no boosts or spin) has, according to our numerical results, a
net linear momentum in the opposite direction from that implied by the
first term of Eq.~(\ref{Eq:PvecBS-A}). 

\section{Analytic calculation of total angular momentum}
\label{AppB}
For conformally flat initial data on CMC hypersurfaces, the expression in
Eq.~(\ref{Eq:AngMom}) for the angular momentum at $\dot{\ScriPlus}$ 
is just an angular integral over the conformal extrinsic curvature
$\tilde{A}_{\ell}^{~k}$, independent of the conformal factor $\Omega$.
The generalized Bowen-York expression for $\tilde{A}_{\ell}^{~k}$, as
given in Eq.~(\ref{Eq:GeneralizedBY}), is a simple sum over terms for
each black hole, each of which contributes independently to the
angular momentum.  Therefore, we can consider each black hole and each
term separately.  For a given
black hole, we can rotate the Cartesian coordinates in the conformal
flat space to locate the black hole with some displacement $\mathbf D$
from the origin to put it on the positive $z$-axis.  The
unit vector $\mathbf N$ normal to this sphere at $R = R_+$, the unit 
vector $\mathbf n$ directed away from the black hole, and the coordinate
distance $r$ from the black hole to a point on $\dot{\ScriPlus}$ are
given in Appendix~\ref{AppA}.  Also note Eq.~(A3) for the scalar and vector
products of $\mathbf N$ and $\mathbf n$.  Define $q \equiv D/{R_+}$
and $x \equiv \cos \theta$.

The contribution from the first term in Eq.~(\ref{Eq:GeneralizedBY}) 
is 
\begin{equation}
- \frac{C} {{8\pi }}\oint {\sin \theta d\theta d\phi \left( {\frac{{R_
          + }} {r}} \right)^3 \left[ {3\left( {\mathbf N \times \mathbf n}
      \right)\left( {\mathbf n \cdot \mathbf N} \right) - \left( {\mathbf N
          \times \mathbf N} \right)} \right]} , \nonumber
\end{equation}
which is identically zero since the integrand has only odd powers 
of $\sin \phi$ and $\cos \phi$.

The ``spin'' term contribution to the total angular momentum is
\begin{eqnarray}
  \mathbf J_S &=& \frac{3} {{8\pi }}\oint {\sin \theta d\theta d\phi \left(
      {\frac{{R_ + }}
        {r}} \right)^3 \left[ {\mathbf N \times \left( {\mathbf S \times
            \mathbf n} \right)\left( {\mathbf n \cdot \mathbf N} \right) + \mathbf
        N \cdot \left( {\mathbf S \times \mathbf n} \right)\left( {\mathbf N
            \times \mathbf n} \right)} \right]}  \nonumber \\
  &=& \frac{3} {4}\int\limits_{ - 1}^1 {\frac{{dx}}
    {{\left( {1 + q^2  - 2qx} \right)^{3/2} }}\left\langle {\left( {\mathbf N \cdot \mathbf n} \right)^2 \mathbf S - \left( {\mathbf N \cdot \mathbf S} \right)\left( {\mathbf n \cdot \mathbf N} \right)\mathbf n + \mathbf N \cdot \left( {\mathbf S \times \mathbf n} \right)\left( {\mathbf N \times \mathbf n} \right)} \right\rangle _\phi  }. 
\end{eqnarray}
Consider separately the components of $\mathbf J$ along and perpendicular
to the displacement of the black hole.  The component parallel to
$\mathbf D$ is
\begin{eqnarray}
  J_S^z &=& \frac{3} {4}S^z \left[ {\int\limits_{ - 1}^1
      {dx\frac{{\left( {1 - qx} \right)^2 }} {{\left( {1 + q^2 - 2qx}
            \right)^{5/2} }}} - \int\limits_{ - 1}^1 {dx\frac{{x\left(
              {1 - qx} \right)\left( {x - q} \right)}}
        {{\left( {1 + q^2  - 2qx} \right)^{5/2} }}} } \right]
  \nonumber \\
  &=& \frac{3} {4}S^z \left[ {\frac{2} {3}\frac{{3 - 2q^2 }} {{1 - q^2
        }} - \frac{2} {3}\frac{1}
      {{1 - q^2 }}} \right] = S^z .
\end{eqnarray}
These integrals (and those below) were done in {\tt Mathematica}.  The
average over $\phi$ kills the contributions of $S^x$ and $S^y$ to $
J_S^z$.

In the expression for the perpendicular component $J_S^x$, there are 
non-trivial contributions from all three terms in the average over $\phi$, 
with only terms proportional to $S^x$ surviving,
\begin{eqnarray}
  J_S^x &=& \frac{3} {4}S^x \left[ {\int\limits_{ - 1}^1
      {dx\frac{{\left( {1 - qx} \right)^2 }} {{\left( {1 + q^2 - 2qx}
            \right)^{5/2} }}} - \frac{1} {2}\int\limits_{ - 1}^1
      {dx\frac{{\left( {1 - x^2 } \right)\left( {1 - qx} \right)}}
        {{\left( {1 + q^2 - 2qx} \right)^{5/2} }} - \frac{1} {2}q^2
        \int\limits_{ - 1}^1 {dx\frac{{\left( {1 - x^2 } \right)}}
          {{\left( {1 + q^2  - 2qx} \right)^{5/2} }}} } } \right]
  \nonumber \\
  &=& \frac{3} {4}S^x \left[ {\frac{2} {3}\frac{{3 - 2q^2 }} {{1 - q^2
        }} - \frac{2} {3} - \frac{2} {3}\frac{{q^2 }}
      {{1 - q^2 }}} \right] = S^x .
\end{eqnarray}
Similarly, $J_S^y = S^y $.  The spin is unmodified by any factors depending 
on the ratio $q =D/R_ + $.  

Now consider first the normal boost terms in Eq.~(\ref{Eq:GeneralizedBY}) 
arising from the boost vector $\mathbf P$,
\begin{eqnarray}
  \mathbf J_P &=& \frac{{3R_ + }} {{16\pi }}\oint {\sin \theta d\theta
    d\phi \left( {\frac{{R_ + }}
        {r}} \right)^2 \left[ {\left( {\mathbf N \times \mathbf P}
        \right)\left( {\mathbf n \cdot \mathbf N} \right) + \left( {\mathbf N
            \cdot \mathbf P} \right)\left( {\mathbf N \times \mathbf n} \right)
        + \left( {\mathbf n \cdot \mathbf P} \right)\left( {\mathbf n \cdot
            \mathbf N} \right)\left( {\mathbf N \times \mathbf n} \right)}
    \right]}  \nonumber \\
  &=& \frac{3} {8}R_ + \int\limits_{ - 1}^1 {\frac{{dx}}
    {{\left( {1 + q^2  - 2qx} \right)}}\left\langle {\left( {\mathbf N \times \mathbf P} \right)\left( {\mathbf n \cdot \mathbf N} \right) + \left( {\mathbf N \cdot \mathbf P} \right)\left( {\mathbf N \times \mathbf n} \right) + \left( {\mathbf n \cdot \mathbf P} \right)\left( {\mathbf n \cdot \mathbf N} \right)\left( {\mathbf N \times \mathbf n} \right)} \right\rangle _\phi  } .
\end{eqnarray}
Again, the average eliminates all terms containing odd powers of $\sin
\phi$ or $\cos \phi$.  We see immediately that only $P^y$ contributes to 
$J_P^x$ and only $P^x$ contributes to  and $J_P^y$.  In particular,
\begin{eqnarray}
  J_P^x &=& - \frac{3} {8}R_ + P^y \left[ {\int\limits_{ - 1}^1
      {dx\frac{{x\left( {1 - qx} \right)}} {{\left( {1 + q^2 - 2qx}
            \right)^{3/2} }} + \frac{q} {2}\int\limits_{ - 1}^1
        {dx\frac{{\left( {1 - x^2 } \right)}} {{\left( {1 + q^2 - 2qx}
              \right)^{3/2} }} + \frac{q} {2}\int\limits_{ - 1}^1
          {dx\frac{{\left( {1 - x^2 } \right)\left( {1 - qx} \right)}}
            {{\left( {1 + q^2  - 2qx} \right)^{5/2} }}} } } } \right]
  \nonumber \\
  &=& - \frac{3} {8}R_ + P^y \left[ {\frac{4} {3}q + \frac{2} {3}q +
      \frac{2}
      {3}q} \right] =  - DP^y  = \left( {\mathbf D \times \mathbf P} \right)^x ,
\end{eqnarray}
and similarly for $J_P^y $.  

The other boost vector, $\mathbf Q$, 
potentially could contribute to the angular momentum, because 
even though its contribution to the extrinsic curvature falls off more 
rapidly with coordinate distance from the black hole, $\ScriPlus$ is at 
a finite coordinate radius.  The terms have a similar form to those in 
$\mathbf J_P$, with
\begin{eqnarray}
  J_Q^x &=& \frac{3} {{8R_ + }}Q^y \left[ {\int\limits_{ - 1}^1
      {dx\frac{{x\left( {1 - qx} \right)}} {{\left( {1 + q^2 - 2qx}
            \right)^{5/2} }} + \frac{q} {2}\int\limits_{ - 1}^1
        {dx\frac{{\left( {1 - x^2 } \right)}} {{\left( {1 + q^2 - 2qx}
              \right)^{5/2} }} - \frac{{5q}} {2}\int\limits_{ - 1}^1
          {dx\frac{{\left( {1 - x^2 } \right)\left( {1 - qx} \right)}}
            {{\left( {1 + q^2  - 2qx} \right)^{7/2} }}} } } } \right]
  \nonumber \\
  &=& \frac{3} {{8R_ + }}Q^y \left[ {\frac{4} {3}q\frac{{\left( {2 - q^2
            } \right)}} {{\left( {1 - q^2 } \right)^2 }} + \frac{2}
      {3}\frac{q} {{1 - q^2 }} - \frac{2} {{3}}q\frac{{5 - 3q^2 }}
      {{\left( {1 - q^2 } \right)^2 }}} \right] = 0,
\end{eqnarray}
so in fact, the $\mathbf Q$ boost does not contribute.

Our final result for the angular momentum of a generalized Bowen-York
black hole at an arbitrary displacement $\mathbf D$ from the origin of a CMC
hypersurface is
\begin{equation}
\label{AppEq:AnalAngMom}
\mathbf J = \mathbf S + \mathbf D \times \mathbf P .
\end{equation}
Again, there are 
independent contributions from each black hole in the case of 
multiple black holes.

\end{appendix}

\bibliography{References/References}

\end{document}